\documentclass[
aip,
amsmath,
amssymb,
reprint,
]{revtex4-2}

\usepackage[english]{babel}
\usepackage{graphicx}
\usepackage{float}
\usepackage{bm}
\usepackage{makecell}
\usepackage[utf8]{inputenc}
\usepackage[T1]{fontenc}
\usepackage{mathptmx}
\usepackage{hyperref}
\usepackage{xcolor}
\usepackage{soulutf8}
\usepackage{braket}
\usepackage{siunitx}
\usepackage{amsmath}
\usepackage{caption}
\usepackage{subcaption}

\DeclareMathOperator{\sinc}{sinc}

\makeatletter
\let\@fnsymbol\@fnsymbol@latex
\@booleanfalse\altaffilletter@sw
\makeatother

\newcommand{\orcid}[1]{\href{https://orcid.org/#1}{\includegraphics[width=10pt]{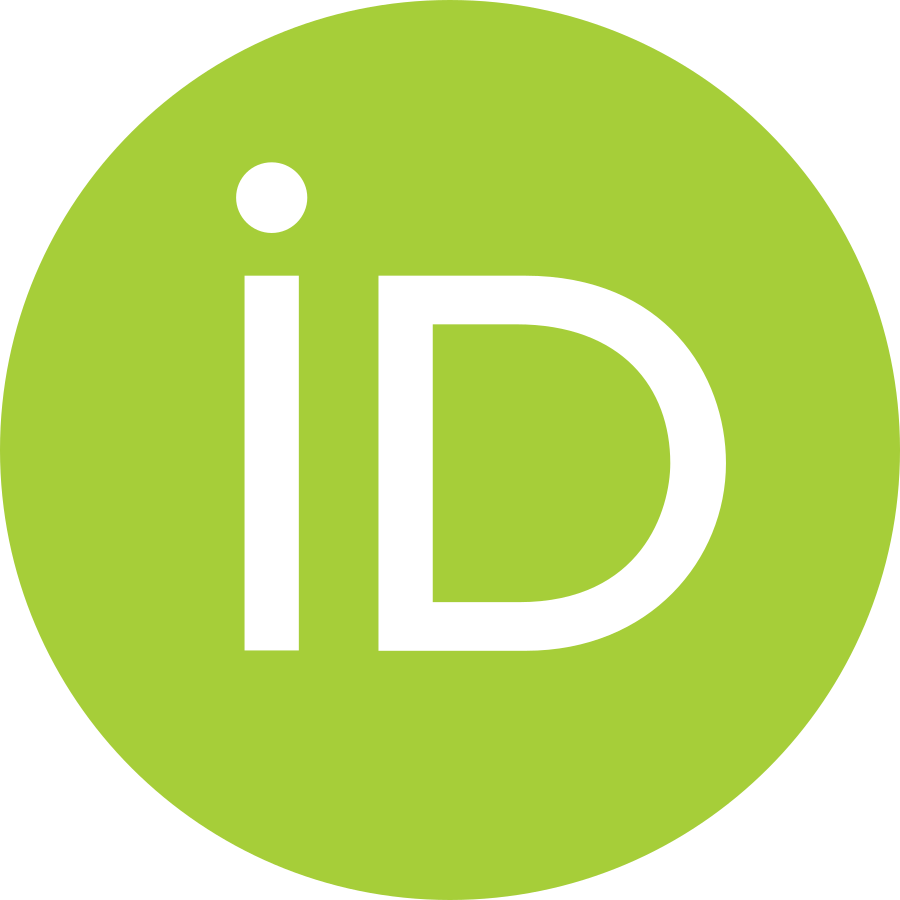}}}

\begin{document}

\title{Photon Pair Source based on PPLN-Waveguides\\ for Entangled Two-Photon Absorption}


\author{Tobias Bernd G\"{a}bler}
\thanks{\href{mailto:tobias.bernd.gaebler@iof.fraunhofer.de}{tobias.bernd.gaebler@iof.fraunhofer.de} \orcid{0000-0003-4874-428X}}
\affiliation{Fraunhofer Institute for Applied Optics and Precision Engineering IOF, Albert-Einstein-Straße 7, 07745 Jena, Germany}
\affiliation{Abbe Center of Photonics, Friedrich Schiller University Jena, Albert-Einstein-Straße 6, 07745 Jena, Germany}

\author{Patrick Hendra}
\affiliation{Fraunhofer Institute for Applied Optics and Precision Engineering IOF, Albert-Einstein-Straße 7, 07745 Jena, Germany}
\affiliation{Abbe Center of Photonics, Friedrich Schiller University Jena, Albert-Einstein-Straße 6, 07745 Jena, Germany}

\author{Nitish Jain}
\affiliation{Fraunhofer Institute for Applied Optics and Precision Engineering IOF, Albert-Einstein-Straße 7, 07745 Jena, Germany}

\author{Markus Gr\"{a}fe}
\affiliation{Fraunhofer Institute for Applied Optics and Precision Engineering IOF, Albert-Einstein-Straße 7, 07745 Jena, Germany}
\affiliation{Institute of Applied Physics, Technical University of Darmstadt, Schloßgartenstraße 7, 64289 Darmstadt, Germany}

\begin{abstract}
    Fluorescence excitation by absorption of entangled photon pairs offers benefits compared to classical imaging techniques, such as the attainment of higher signal levels at low excitation power while simultaneously mitigating phototoxicity. However, current entangled photon pair sources are unreliable for fluorescence detection. In order to address this limitation, there is a need for ultra-bright entangled photon pair sources. Among the potential solutions, sources utilizing nonlinear waveguides emerge as promising candidates to facilitate fluorescence excitation through entangled photons.
    In this paper, a source consisting of a periodically poled lithium niobate waveguide was developed and its key characteristics analysed. To demonstrate its suitability as key component for imaging experiments, the entangled two-photon absorption behavior of CdSe/ZnS quantum dot solutions was experimentally investigated. 
\end{abstract} 
\maketitle

\section{Introduction}
\label{sec:Introduction}

Fluorescence caused by the absorption of entangled photon pairs (eTPA) can enable a new generation of micro\-scopes, which combine benefits of conventional one- and two-photon fluorescence imaging~\cite{Mukamel2020,Raymer2021a,Gilaberte2019}. The absorption rate of entangled photon pairs depends linearly on the excitation power~\cite{Fei1997} and, thus, reducing photobleaching due to the higher signal compared to classical two-photon excitation in the low light regime. Naturally, two-photon absorption processes operate at larger wavelengths compared to single-photon absorption. This leads to less scattering inside the tissue which results in deeper penetration depths~\cite{Lissandrin2004} which is also applicable for eTPA.\\
However, a practical application of fluorescence imaging by eTPA is not possible with state-of-the-art photon pair sources since the power of generated entangled photon pairs is too low. For this reason, an explicit detection of fluorescence photons, which are traced back to the entangled photon pair absorption process, becomes difficult. Previous studies tried to solve this issue by using pulsed lasers or complex fluorescence molecules~\cite{Varnavski2017,Villabona-Monsalve2017,Parzuchowski2020,Villabona-Monsalve2020}. Both approaches are connected with drawbacks and thus not ideal for a new generation of commercial life science microscopes: Despite the fact that pulsed laser systems can be directly used for fluorescence through classical two-photon absorption (cTPA), their cost remains a notable constraint. On the other hand, injecting complex molecules might change the structure of biological probes dramatically.\\
These issues, such as the requirement for high-power laser systems to overcome low signal-to-noise ratio and the negative impact of complex fluorescence markers on probes, can potentially be resolved by using highly efficient entangled photon pair sources. Among the common candidates for such light sources, periodically poled nonlinear waveguide crystals represent the best compromise between handling and generation efficiency. Such waveguides generate entangled photon pairs by spontaneous parametric down-conversion (SPDC) ~\cite{Fiorentino2007}.

In this paper, after a brief insight into the fundamental aspects (section~\ref{sec:Fundamentals}) of eTPA, our concept of a bright entangled photon pair source based on periodically poled lithium niobate (PPLN) waveguides and its key characteristics will be shown (section~\ref{sec:EPP Setup}). The practical application of this source for entangled two-photon fluorescence microscopy will be demonstrated by analysing the absorption behavior of different solvents and CdSe/ZnS quantum dots (section~\ref{sec:Absorption Measurements}). Crucial parts of our experimental investigations are described and possible weaknesses addressed.\\

\section{Fundamentals}
\label{sec:Fundamentals}

The crucial point of every fluorescence imaging technique is the accurate and efficient excitation of fluorophores. It is directly related to the absorption rate $R$. In the case of two-photon absorption (TPA), the classical and entangled cases differ and lead to linear and quadratic dependency on the incoming photon pair flux $\phi_{\mathrm{Pair}}$, respectively.
\begin{align} 
R_{\mathrm{TPA}} = \sigma_{\mathrm{e}} \phi_{\mathrm{Pair}} + \delta_{\mathrm{c}} \phi_{\mathrm{Pair}}^2, \label{eq:AbsorptionRate}
\end{align}
$\sigma_{\mathrm{e}}$ and $\delta_{\mathrm{c}}$ represent the entangled and classical two-photon absorption cross sections. Both are related by 
\begin{align} 
\sigma_{\mathrm{e}} \propto \frac{\delta_{\mathrm{c}}}{T_{\mathrm{e}} A_{\mathrm{e}}} \label{eq:EntangledCrossSection}
\end{align}
taking entanglement time $T_{\mathrm{e}}$ and entanglement area $A_{\mathrm{e}}$ ~\cite{Fei1997, Landes2021} into consideration. These quantities represent the temporal and spatial correlation of two photons within the entangled pair, which are specified by the generation process. Meanwhile, the dispersion along the optical path introduces additional temporal and spatial mismatches, which reduce absorption rates and therefore the total fluorescence signal.\\
However, an efficient setup designed for entangled two-photon fluorescence microscopy must be adjusted so that $T_{\mathrm{e}}$ and $A_{\mathrm{e}}$ become as small as possible to maximize $\sigma_{\mathrm{e}}$. Since the spatial mismatch is mainly determined by focusing the entangled photon beam (with wavelength $\lambda$) into the sample, $A_{\mathrm{e}}$ can be approximated to the spot size given by the Airy disc diameter $d_{\mathrm{Airy}}$.\\
\begin{align} 
A_{\mathrm{e}} \approx \frac{\pi}{4}d_{\mathrm{Airy}}^2 =\frac{\pi}{4}\left(1.22\frac{\lambda}{NA}\right)^2 \label{eq:EntanglementArea}
\end{align}
In contrast to $A_{\mathrm{e}}$, entanglement time $T_{\mathrm{e}}$ is already studied in more detail 
 ~\cite{Fei1997,Parzuchowski2020,Tabakaev2021,Corona-Aquino2022,Landes2021}. Considering the SPDC process in a nonlinear crystal of length $L$, $T_{\mathrm{e}}$ is determined by 
\begin{align} 
T_{\mathrm{e}} = \frac{L}{2}\cdot \left\vert\frac{1}{v_{\mathrm{gr}}^{\mathrm{s}}}-\frac{1}{v_{\mathrm{gr}}^{\mathrm{i}}}\right\vert. \label{eq:EntanglementTime}
\end{align}
$v_{\mathrm{gr}}^{\mathrm{s,i}}=\partial \omega_{\mathrm{s,i}} / \partial k_{\mathrm{s,i}}$ represent the group velocities of both photons of an entangled photon pair inside the crystal, where $\mathrm{s}$ and $\mathrm{i}$ are traditionally denoted as signal and idler photon. The exact values of these group velocities depend on the energy and transversal phase matching conditions 
\begin{align} 
\omega_{\mathrm{p}} &= \omega_{\mathrm{s}} + \omega_{\mathrm{i}} \label{eq:energy conservation}\\ \Delta k &= k_{\mathrm{p}} - k_{\mathrm{s}} - k_{\mathrm{i}} - \frac{2\pi}{\Lambda}\label{eq:phase matching}
\end{align}
during the SPDC process, where $\omega_{\mathrm{p}}$ denotes the center frequency of the pump photon ~\cite{Fei1997, Mertenskoetter2021}. The transverse wave vectors $k_{\mathrm{p,s,i}}=n_{\mathrm{p,s,i}}\omega_{\mathrm{p,s,i}} / c_0$ depend on the frequency- and temperature-dependent refractive indices $n_{\mathrm{p,s,i}}\left(\omega_{\mathrm{p,s,i}},\vartheta\right)$ of the nonlinear crystal ~\cite{Gayer2008}. Temperature independent phase matching for the SPDC process is achievable in several materials, such as barium borate (BBO) or potassium titanyl phosphate (KTP), for specific wavelength and polarization combinations of pump, signal and idler ~\cite{Karan2020}. Furthermore, alternating poled domains can be fabricated in ferroelectric nonlinear materials, like in lithium niobate (LN) or KTP, which enable efficient SPDC with a wide range of wavelength combinations of signal and idler photons. In this case, $\Lambda$ represents the length of these periodically poled domains inside such crystals and causes the additional phase term as given in eq. \eqref{eq:phase matching} as well as a strong dependence on the crystal temperature ~\cite{Fiorentino2007}.\\
The spectrum of the generated photon pairs is given by the two-photon wave function
\begin{align} 
\ket{\psi_{\mathrm{SPDC}}} &\propto \int f\left(\omega_{\mathrm{s}},\omega_{\mathrm{i}}\right)d\omega_{\mathrm{s}}d\omega_{\mathrm{i}} \ket{1}_{\mathrm{s}}\ket{1}_{\mathrm{i}} \label{eq:SPDCWaveFunction} 
\end{align} 
where $f\left(\omega_{\mathrm{s}},\omega_{\mathrm{i}}\right)$ is called joint spectral amplitude (JSA) ~\cite{Kim2021,Kumar2019}.
\begin{align}
f\left(\omega_{\mathrm{s}},\omega_{\mathrm{i}}\right)&=\alpha\left(\omega_{\mathrm{s}},\omega_{\mathrm{i}}\right)\varphi\left(\omega_{\mathrm{s}},\omega_{\mathrm{i}}\right)\label{eq:JSA} 
\end{align} 
It consists of pump envelope $\alpha$ with pump wavelength variance $\sigma_{\mathrm{p}}$ and phase matching function $\varphi$ with crystal length $L$.
\begin{align}
\alpha\left(\omega_{\mathrm{s}},\omega_{\mathrm{i}}\right) &= \exp\left(-\frac{\left(\omega_{\mathrm{s}}+\omega_{\mathrm{i}}-\omega_{\mathrm{p}}\right)^2}{2\sigma_{\mathrm{p}}^2}\right) \label{eq:PumpEnvelope} \\
\varphi\left(\omega_{\mathrm{s}},\omega_{\mathrm{i}}\right) &=\sinc\left(\frac{\Delta k\left(\omega_{\mathrm{s}},\omega_{\mathrm{i}}\right) L}{2}\right)\label{eq:PhaseMatchingFunction} 
\end{align} 
In summary, the successful occurrence of eTPA depends on several setup-related parameters. As also was done in other publications, for example in references ~\cite{Leon-Montiel2019,Mertenskoetter2021}, we conducted an extensive characterisation of our photon pair source in section \ref{sec:EPP Setup} to evaluate important parameters before its usage for eTPA experiments (section \ref{sec:Absorption Measurements}).

\section{Entangled Photon Pair Source}
\label{sec:EPP Setup}

\begin{figure*}
 \centering
 \includegraphics[width=\textwidth]{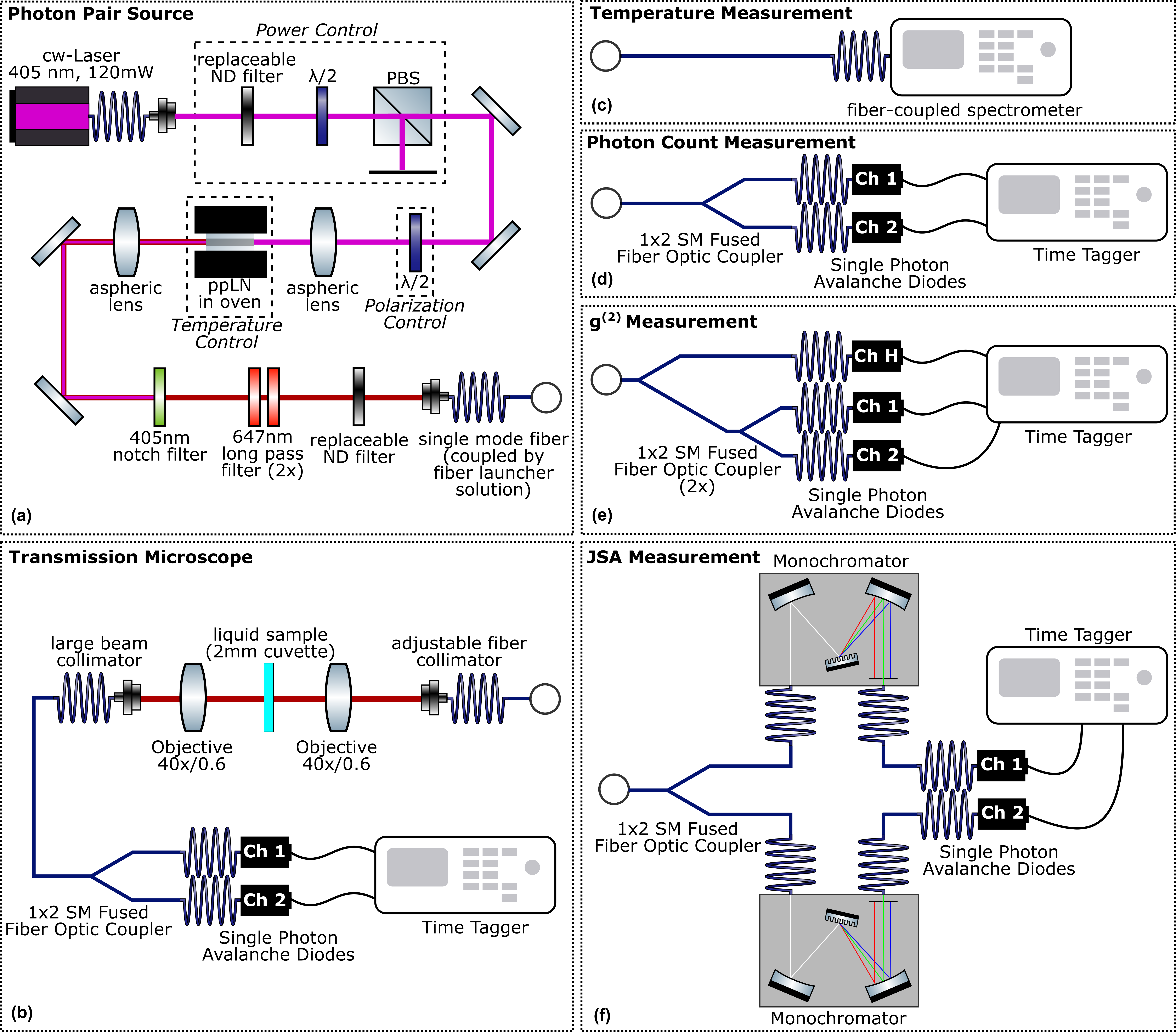}
 \caption{Sketch of the modular experimental setup. The properties and applicability for eTPA of the photon pair source (a) are analysed by different measurement setups (b)-(f).}
 \label{fig:Setup}
\end{figure*}

A complete sketch of our experimental setup is shown in fig. \ref{fig:Setup}. To pave the way towards low-cost and sensitive two-photon fluorescence microscopy, our setup bases on a modular and fiber-coupled design which enables the possibility of using the photon pair source (fig. \ref{fig:Setup}a) with any commercial or scientific two-photon microscope as well as for applications in photonic quantum information technologies.

The generation of time-entangled photon pairs is driven by a fiber-coupled $\SI{405}{\nano\metre}$ CW-laser (Toptica iBeam-Smart-405-S-HP). The light beam of this laser is collimated by a fiber coupler (Schäfter + Kirchhoff 60FC-0-M12-33). With a removable neutral density filter, a half-wave plate and a polarization beam splitter, the power of the pump beam can be controlled. The second half-wave plate is used to set the polarization of the pump. After two tip-tilt-mirrors, the beam is focused by an aspheric lens (Thorlabs C220TMD-A) onto a waveguide of ppLN (AdVR) with a length of $L=\SI{20}{\milli \metre}$, a poling period of $\Lambda=\SI{2.72}{\micro\metre}$ and an anti-reflection coating in the ranges of $\SI[separate-uncertainty = true]{405(5)}{\nano\metre}$ and $\SI[separate-uncertainty = true]{810(50)}{\nano\metre}$. This nonlinear waveguide enables the efficient generation of entangled photons with a center frequency of $\SI{810}{\nano\metre}$ by type-0 SPDC. For tunability of phase matching and, thus, the wavelength combinations of generated entangled photons, the nonlinear waveguide chip is placed in an oven (Covesion PV20).\\
Following the waveguide, the light beam, consisting of pump and entangled photons, is collimated by another aspheric lens (Thorlabs C220TMD-B) and filtered by a combination of a notch filter (Thorlabs NF405-13) and two identical long pass filters (Semrock BLP01-647R-25) to remove pump photons, thus preventing single-photon excitations of the probed fluorescence dyes. Replaceable neutral density filters allow the consideration of the classical absorption behavior as explained later in section \ref{sec:Absorption Measurements}. The beam is guided by two mirrors to a fiber launch solution for coupling into single mode fibers.\\

\subsection{Temperature Dependency}
\label{sec:Temperature Dependency}

\begin{figure}
 \centering
 \includegraphics[width=0.9\columnwidth]{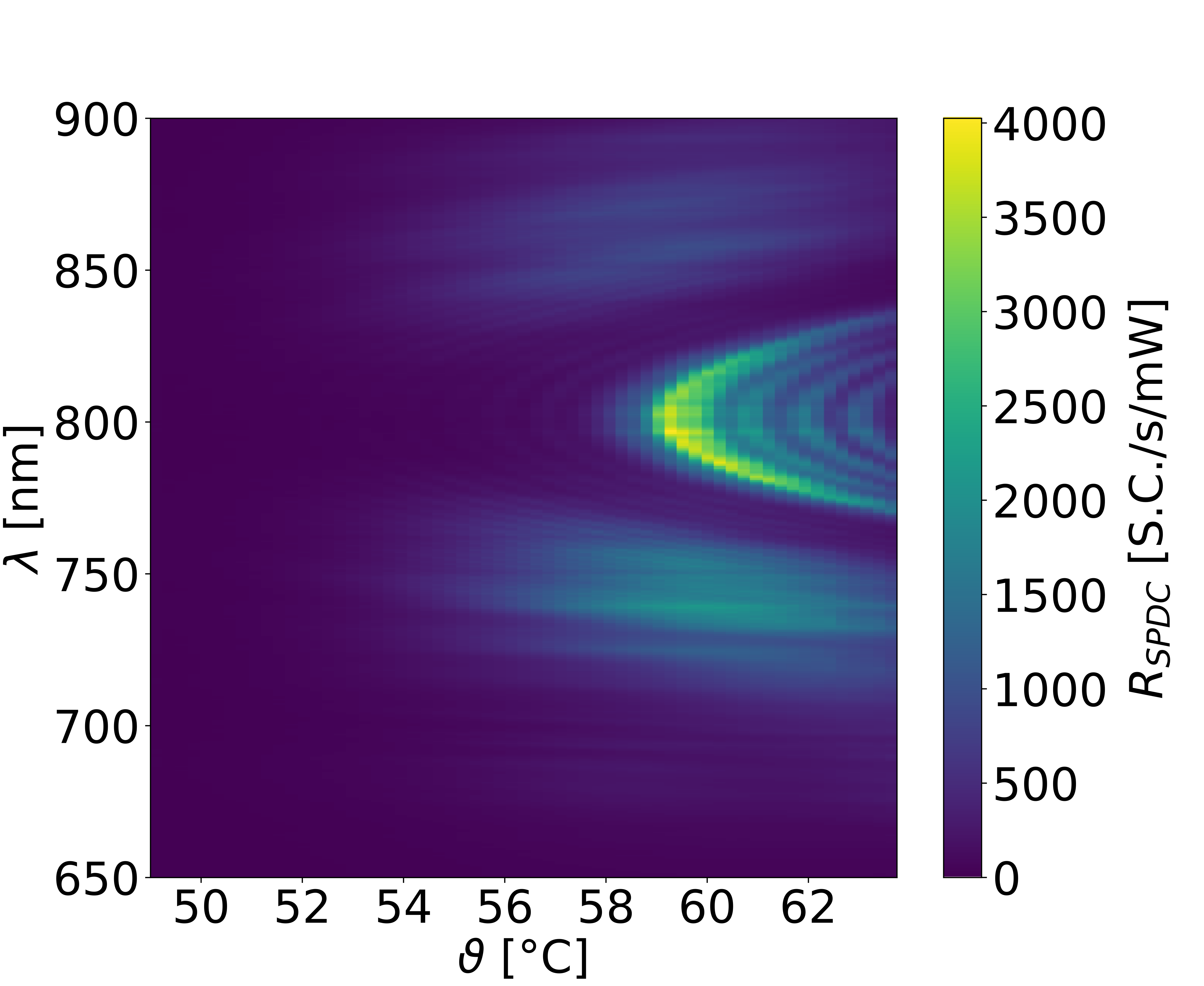}
 \caption{Measured single photon count rate $R_\mathrm{SPDC}$ for different waveguide temperatures $\vartheta$. The degenerated case is given at $\vartheta \approx \SI{59.40}{\celsius}$}
 \label{fig:Temperature Map}
\end{figure}

The dependency of the generated entangled photons on the waveguide temperature was investigated by measuring the single photon count rate $R_\mathrm{SPDC}$ with a highly sensitive spectrometer Ocean Insight QE Pro (fig. $\ref{fig:Setup}$c) for different oven temperatures. To collect as many photons as possible that pass the single-mode fiber, a multi-mode fiber (Thorlabs M134L01) is used to connect the spectrometer.\\
Fig. \ref{fig:Temperature Map} shows the dependency of entangled photon emission spectra on the wave\-guide temperature. The parabolic shape results from the conservation of energy and phase matching conditions \eqref{eq:energy conservation}-\eqref{eq:phase matching}. Because of the usage of a nonlinear waveguide and a single-mode fiber, the wavelength bandwidths of signal and idler photons are narrower in comparison to the usage of bulk crystals. This enables a more accurate setting of various wavelength combinations for spectroscopic applications.\\
The highest count of emitted photons is reached at $\vartheta \approx \SI{59.5}{\celsius}$ which correspond to the case of degenerated SPDC ($\omega_{\mathrm{s,i}}=\omega_{\mathrm{p}}/2$). For all measurements described in the following, the waveguide temperature was fixed on this value.\\

\subsection{Pair Generation Efficiency}
\label{sec:Pair Generation Efficiency}

\begin{figure*}
    \centering
    \begin{subfigure}{0.48\textwidth}
        \centering
        \includegraphics[width=\textwidth]{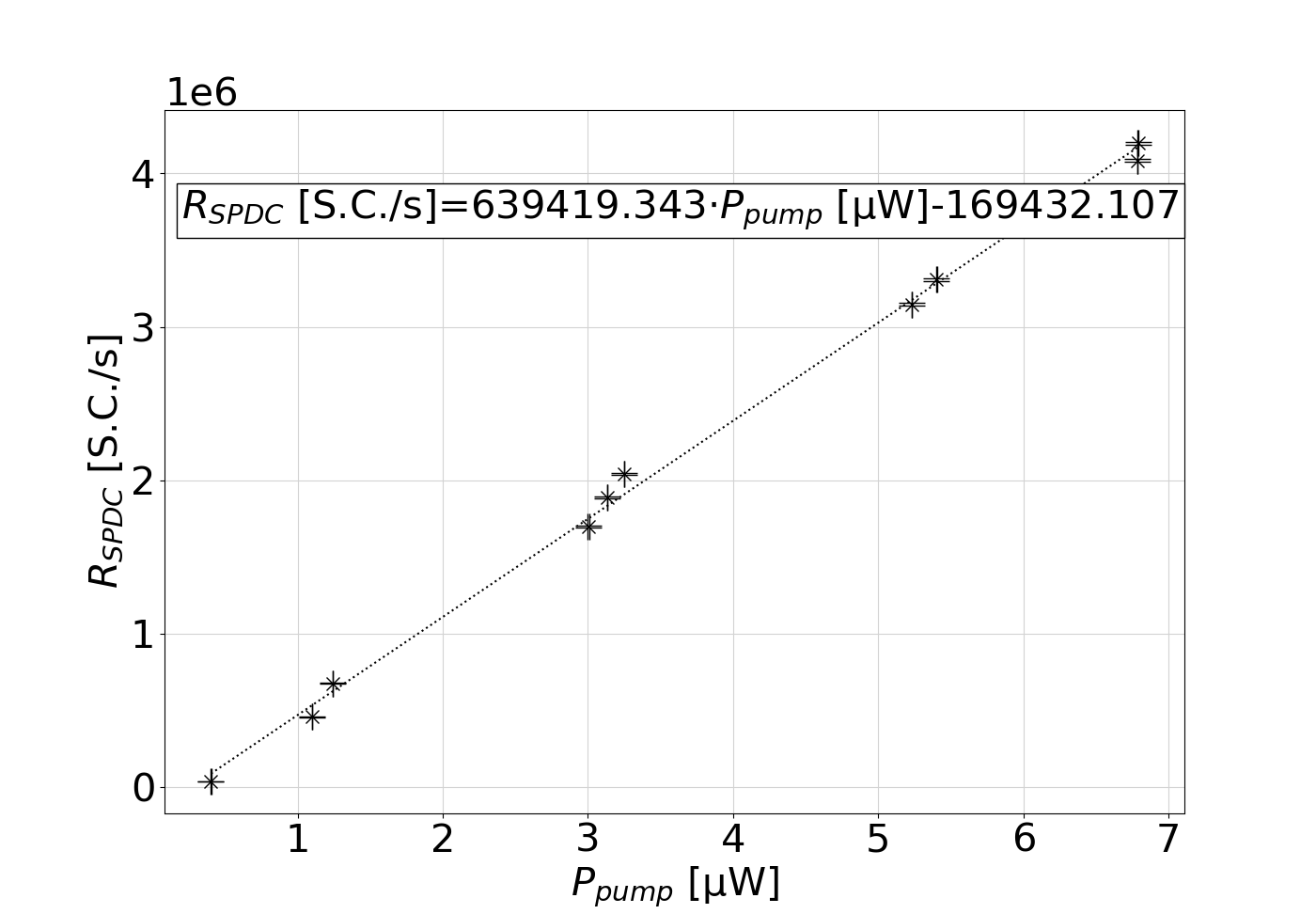}
        \caption{Single photon generation}
        \label{fig:RSPDC Sing}
    \end{subfigure}
    \hfill
    \begin{subfigure}{0.48\textwidth}
        \centering
        \includegraphics[width=\textwidth]{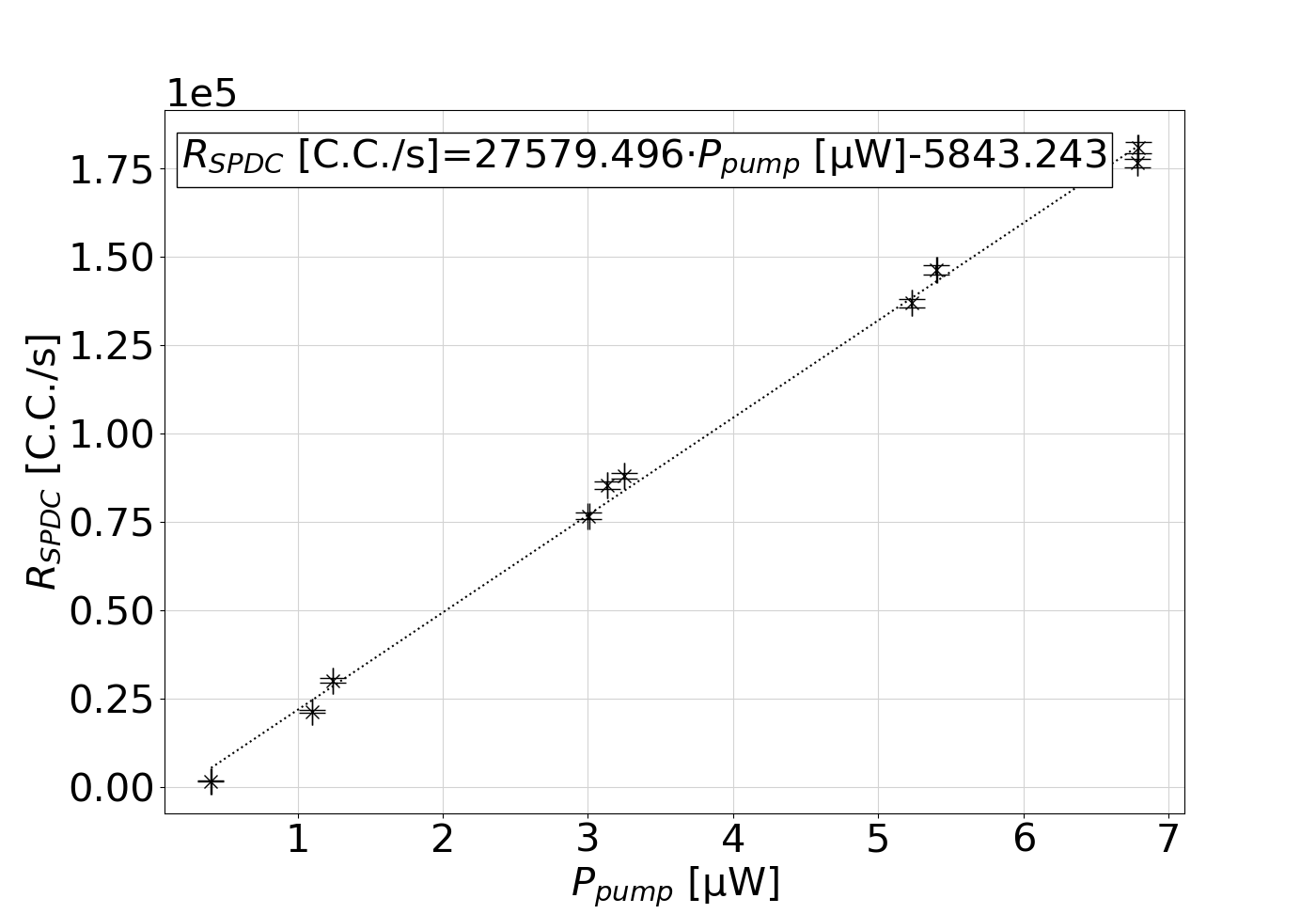}
        \caption{Photon pair generation}
        \label{fig:RSPDC Coin}
    \end{subfigure}
 \caption{Emitted photon counts $R_{\mathrm{SPDC}}$ of the entangled photon pair source for variable pump power $P_{\mathrm{pump}}$ onto the nonlinear waveguide. \ref{fig:RSPDC Sing} illustrates single photon rate, \ref{fig:RSPDC Coin} the coincidence count rate. Losses due to fiber coupling and detection efficiency are not included.}
 \label{fig:Source Counts}
\end{figure*}

\begin{figure}
 \centering
 \includegraphics[width=\columnwidth]{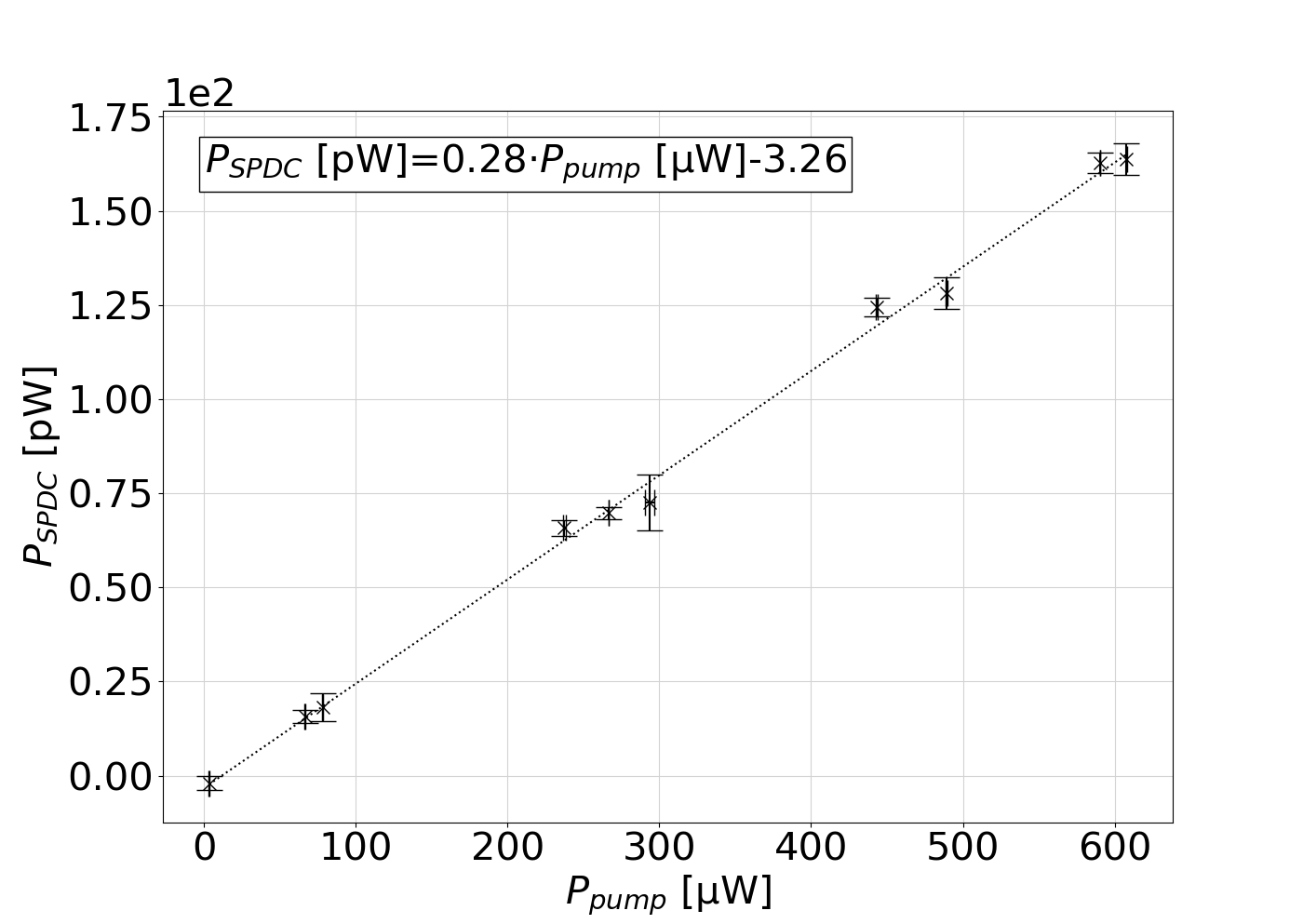}
 \caption{Measured power $P_{\mathrm{SPDC}}$ of down-converted photons at approximately $\SI{810}{\nano\metre}$ for different pump levels $P_{\mathrm{pump}}$.}
 \label{fig:Power calibration}
\end{figure}

Using the setup shown in fig. \ref{fig:Setup}d, the single and coincidence count rate are measured by using a single mode fused fiber optic coupler (Thorlabs TN808R5F1), two single photon avalanche diodes (Excelitas SPCM-800-42-FC) and a time tagger (QuTools quTAG standard). The pump power $P_{\mathrm{pump}}$ was limited to $\SI{7}{\micro\watt}$ in order not to overload the detectors. The photon counting was executed with an integration time of $\SI{100}{\milli\second}$ and a coincidence window of $\SI{1}{\nano\second}$. \\
All data points shown in the following figures were corrected by dark and accidental count rates, but not by losses caused by the experimental setup. However, the detection efficiency of the avalanche diodes is $\eta_{\mathrm{det}}\approx\SI{60}{\percent}$ for $\SI{800}{\nano\metre}$. We also estimate a single mode fiber coupling efficiency of $\eta_{\mathrm{coup}}\approx\SI{90}{\percent}$ \cite{Steinlechner2014,Schwaller2022}. The insertion loss for the used single mode fused fiber optic coupler is given by $\SI{3.7}{\decibel}$, which results in a transmission efficiency through the coupler of $\eta_{\mathrm{inser}}\approx\SI{43}{\percent}$. All these losses finally result in total detection efficiencies of
\begin{align}
\eta_{\mathrm{Singles}} &=\eta_{\mathrm{coup}}\cdot\eta_{\mathrm{inser}}\cdot\eta_{\mathrm{det}}  \approx\SI{23}{\percent} \label{eq:DetectionEfficiencySingles} \\
\eta_{\mathrm{Coin}} &=\eta_{\mathrm{coup}}\cdot\eta^2_{\mathrm{inser}}\cdot\eta^2_{\mathrm{det}}\approx\SI{6}{\percent}\label{eq:DetectionEfficiencyCoin} 
\end{align} 
for single and coincidence count rates.\\
Fig. \ref{fig:Source Counts} shows the detected single counts ($\mathrm{S.C.}$) and coincidence counts ($\mathrm{C.C.}$) generated by the photon pair source. The regression analysis results in $R_{\mathrm{SPDC}}^{\mathrm{Singles}}\approx\SI{6.4e5}{\frac{\mathrm{S.C./}\second}{\micro\watt_\mathrm{pump}}}$ and $R_{\mathrm{SPDC}}^{\mathrm{Coin}}\approx\SI{2.7e4}{\frac{\mathrm{C.C./}\second}{\micro\watt_\mathrm{pump}}}$. This leads to a generation efficiency of $\eta_{\mathrm{SPDC}}\approx\SI{0.28}{\frac{\nano\watt_{\mathrm{SPDC}}}{\milli\watt_{\mathrm{pump}}}}$ as illustrated in fig. \ref{fig:Power calibration}. Taking $\eta_{\mathrm{Coin}}$ into account, our source should be able to achieve emitted photon pair rates of $R_{\mathrm{SPDC}}^{\mathrm{Coin}}\approx\SI{4.5e5}{\frac{\mathrm{C.C./}\second}{\micro\watt_\mathrm{pump}}}$ in an ideal, free-space configuration.\\

\subsection{Second-Order Intensity Correlation}
\label{sec:g(2)-measurement}

\begin{figure}
 \centering
 \includegraphics[width=\columnwidth]{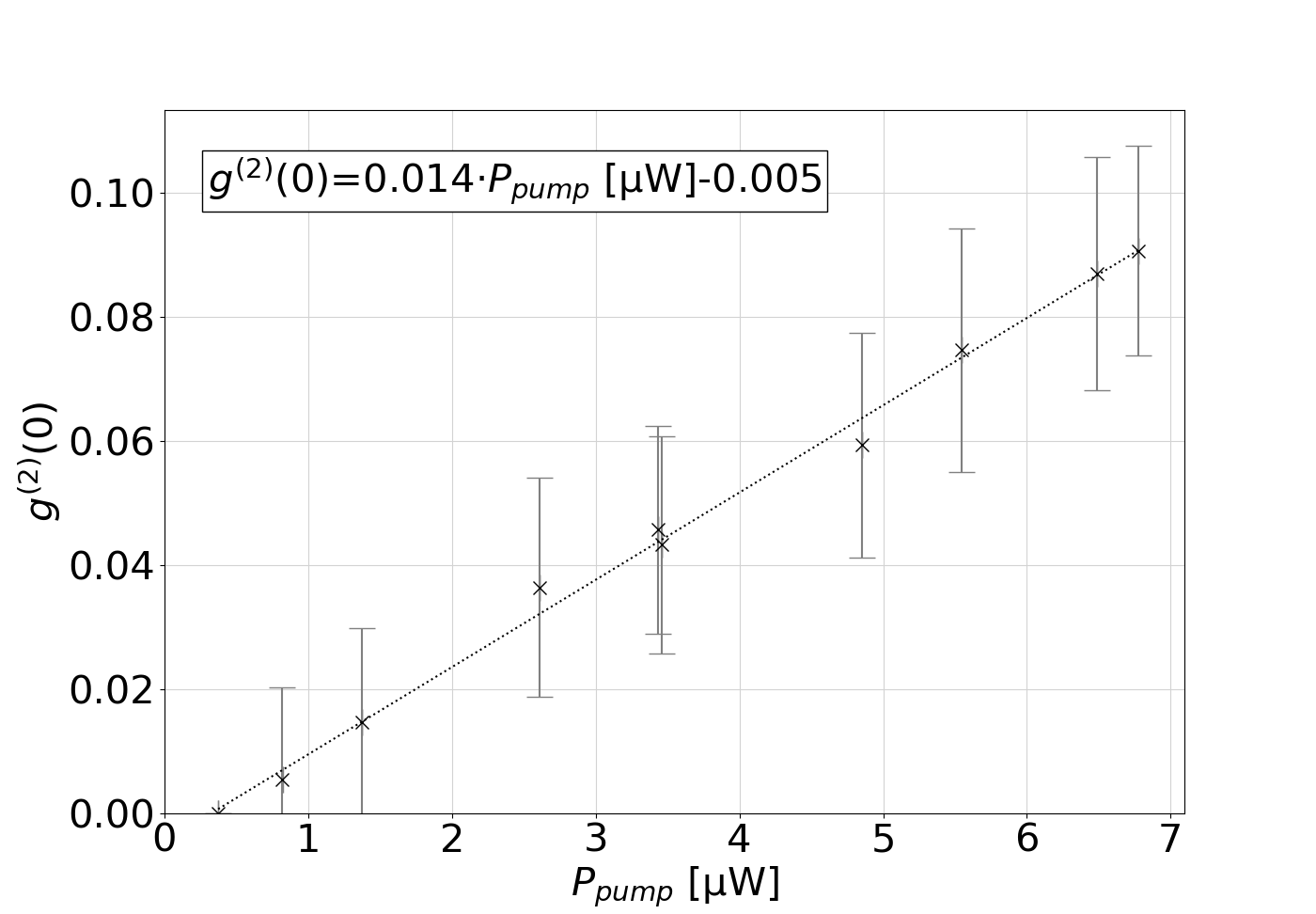}
 \caption{Second-order intensity correlation $g^{\left(2\right)}\left(0\right)$ for various pump power $P_{\mathrm{pump}}$.}
 \label{fig:g(2)-source}
\end{figure}

To evaluate the quality of the source as an emitter of entangled photons, its heralded second-order intensity correlation $g^{\left(2\right)}\left(0\right)$ has to be considered. For this purpose, the setup shown in fig. \ref{fig:Setup}e consisting of two fused single mode fiber (Thorlabs TN808R5F1), three single photon avalanche detectors (Excelitas SPCM-800-42-FC) and a time tagger (QuTools quTAG standard) was used. This setup configuration reduces the amount of coincidence counts caused by accidental events.\\
Performing the measurement of single photon counts $R_{\mathrm{h}}^{\mathrm{Singles}}$ at the heralding channel H and coincidence counts $R_{\mathrm{h,1,2}}^{\mathrm{Coin}}$ between all three channels, $R_{\mathrm{h,1}}^{\mathrm{Coin}}$ between channel H and 1 as well as $R_{\mathrm{h,2}}^{\mathrm{Coin}}$ between channel H and 2 for different pump power, the heralded second-order intensity correlation is given by
\begin{align} 
g^{\left(2\right)}\left(0\right) &= \frac{R_{\mathrm{h}}^{\mathrm{Singles}} \cdot R_{\mathrm{h,1,2}}^{\mathrm{Coin}}}{R_{\mathrm{h,1}}^{\mathrm{Coin}}\cdot R_{\mathrm{h,2}}^{\mathrm{Coin}}}\label{eq:g(2)-equation}. 
\end{align}\\
Fig. \ref{fig:g(2)-source} shows the linear trend of $g^{\left(2\right)}\left(0\right)$ with increasing pump power. Until approximately $P_{\mathrm{pump}}=\SI{70}{\micro\watt}$, $g^{\left(2\right)}\left(0\right)<1$ and, thus, the source emits anti-bunched light until this power level. In our opinion, this is a requirement for the unambiguous observation of eTPA. For $g^{\left(2\right)}\left(0\right)>1$, we assume an increasing amount of classical absorption events caused by photons coming from different pairs. This may avoid a clear distinction of classical and entangled absorption.

\subsection{Entanglement Time}
\label{sec:Entanglement Time}

\begin{figure*}
    \centering
    \begin{subfigure}{0.4\textwidth}
        \centering
        \includegraphics[width=\textwidth]{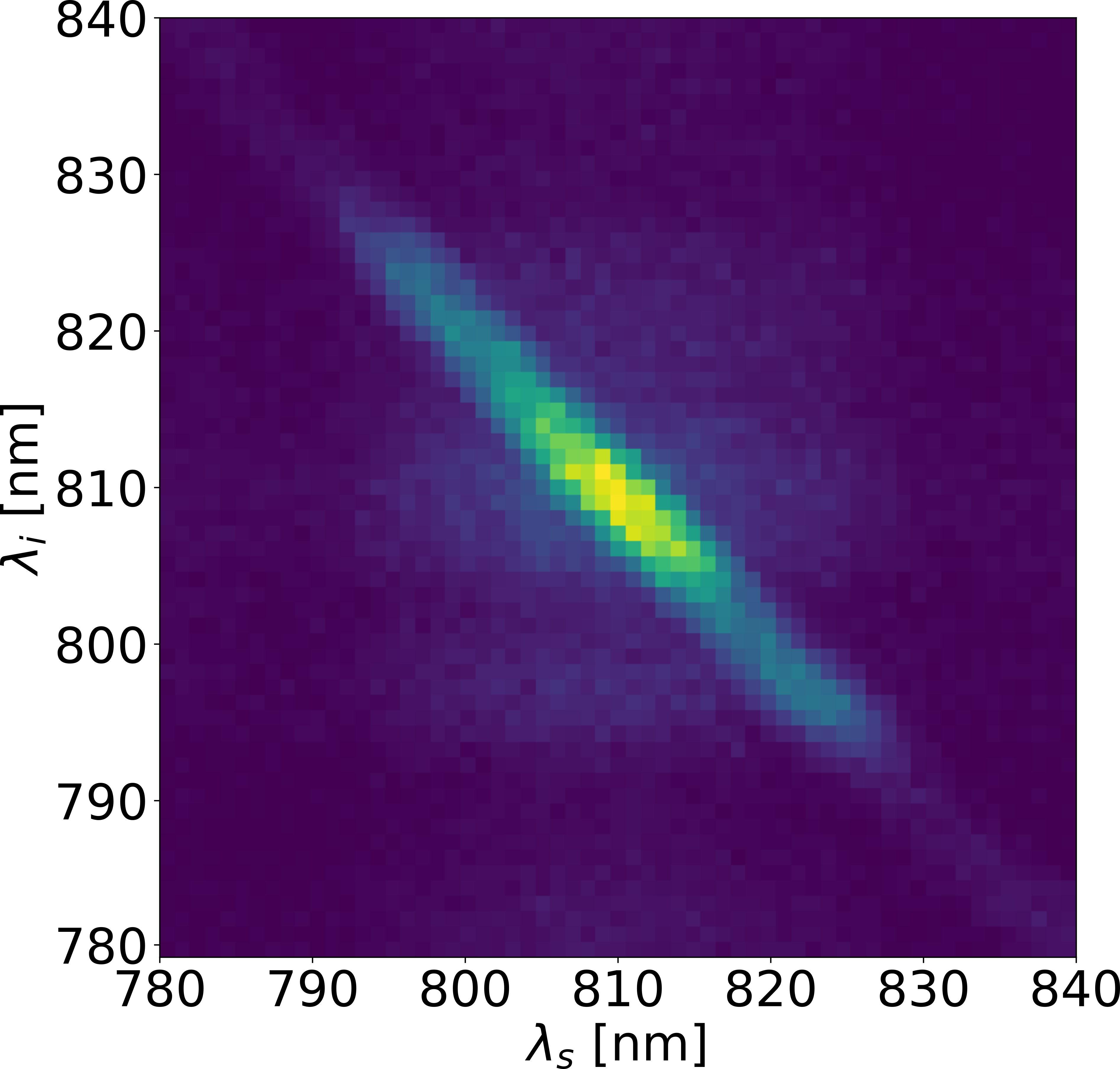}
        \caption{Joint spectral intensity}
        \label{fig:JSI}
    \end{subfigure}
    \hfill
    \begin{subfigure}{0.4\textwidth}
        \centering
        \includegraphics[width=\textwidth]{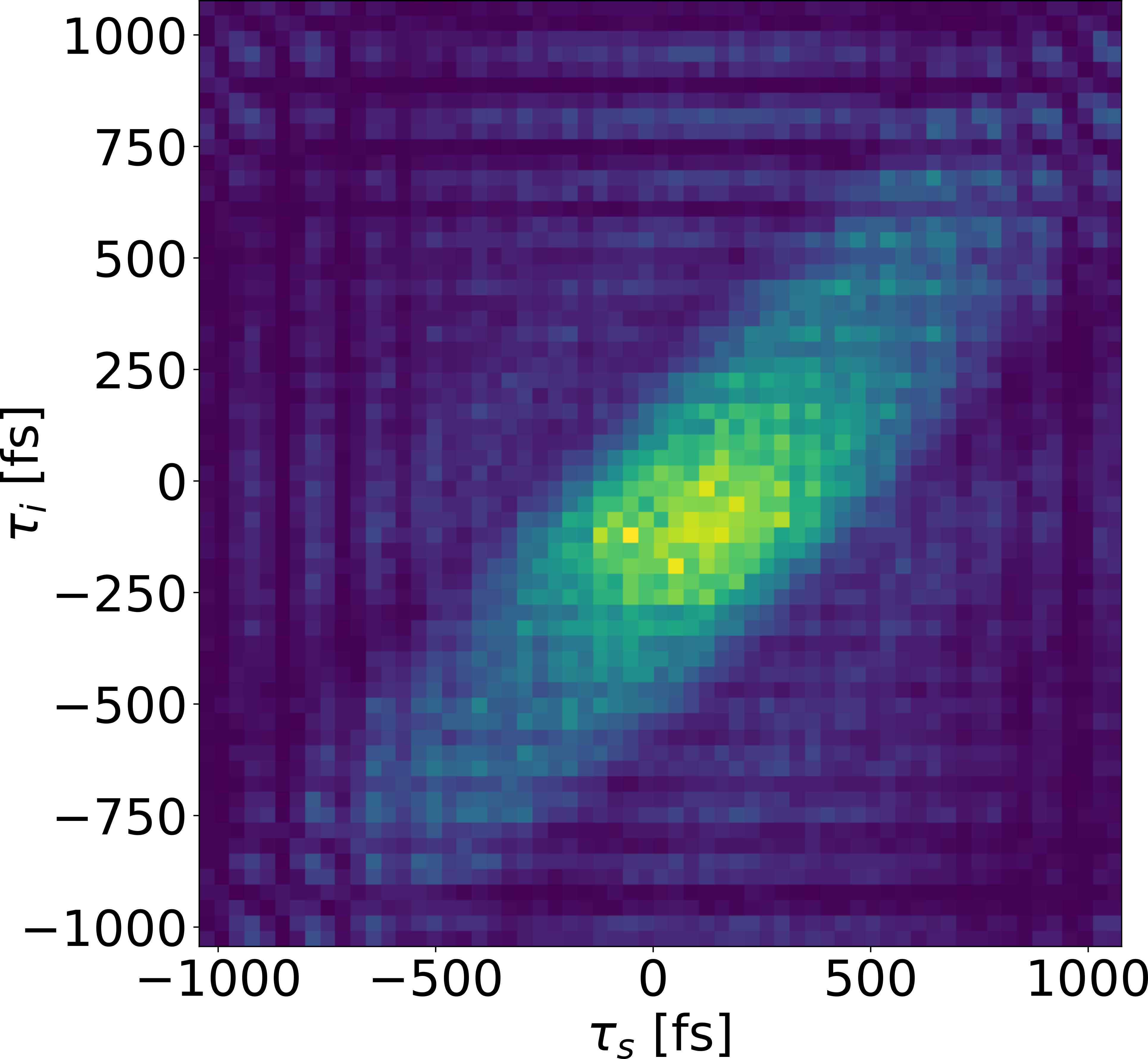}
        \caption{Joint temporal intensity}
        \label{fig:JTI}
    \end{subfigure}
 \caption{Measured joint spectral intensity (\ref{fig:JSI}) and calculated joint temporal intensity (\ref{fig:JTI}). The FWHM along the diagonal of the JTI represents the entanglement time $T_\mathrm{e}\approx\SI{408.6}{\femto\second}$.}
 \label{fig:Entanglement Time}
\end{figure*}

A common way to experimentally determine entanglement time $T_\mathrm{e}$ of an entangled photon pair source is based on the measurement of its joint spectral intensity (JSI) $F\left(\omega_{\mathrm{s}},\omega_{\mathrm{i}}\right)$ which is given by the absolute square of the JSA $f\left(\omega_{\mathrm{s}},\omega_{\mathrm{i}}\right)$ (see eq.~\eqref{eq:JSA}).
\begin{align}
F\left(\omega_{\mathrm{s}},\omega_{\mathrm{i}}\right)&=\left|f\left(\omega_{\mathrm{s}},\omega_{\mathrm{i}}\right)\right|^2
\label{JSI by JSA}
\end{align}
The JSI was recorded by measuring coincidence counts for multiple wavelength combinations selected by two monochromators (lightsource.tech Monochromator-F equipped with optical grids with 1200 Lines/mm and 750~nm blaze) as shown in fig. \ref{fig:Setup}f ~\cite{Zielnicki2018}. For this purpose, the entangled photon pairs are coupled into a fused single mode fiber coupler (Thorlabs TN808R5F1) with a length of $\SI{0.8}{\metre}$ and detected after passing the monochromators and connection fibers (Thorlabs P2-780A-PCSMA-1) with a length of $\SI{1}{\metre}$ by two single photon avalanche detectors (Excelitas SPCM-800-42-FC) and a time tagger (QuTools quTAG standard).\\
Following the approach of reference ~\cite{Parzuchowski2020}, the JSI can be converted into the JSA by
\begin{align}
f\left(\omega_{\mathrm{s}},\omega_{\mathrm{i}}\right)=&\sqrt{F\left(\omega_{\mathrm{s}},\omega_{\mathrm{i}}\right)}\cdot\exp\left(i\frac{\beta}{2}\left(\omega_{\mathrm{s}}-\omega_{\mathrm{p}}\right)^2\right) \notag\\
&\cdot\exp\left(i\frac{\beta}{2}\left(\omega_{\mathrm{i}}-\omega_{\mathrm{p}}\right)^2\right)\label{eq:JSA by JSI} 
\end{align} 
where $\beta\approx\SI{3.3e4}{\femto\second^2}$ represents the group delay dispersion mainly caused by the single mode fibers. The Fourier transformation of the JSA gives the joint temporal amplitude (JTA), whose absolute square leads to the JTI, which allows to extract the entanglement time $T_\mathrm{e}$ ~\cite{Fei1997, Raymer2021a}.
Both, measured JSI and subsequently calculated JTI, are illustrated in fig. \ref{fig:Entanglement Time}. From this we extract an entanglement time of $T_\mathrm{e}\approx\SI{408.6}{\femto\second}$ represented as FWHM along the diagonal axis. In comparison to other studies ~\cite{Parzuchowski2020,Corona-Aquino2022}, this value is substantially larger than reported values due to the dispersion caused by using single mode fibers. An expected value of $T_\mathrm{e}\approx\SI{102}{\femto\second}$ caused only by the group delay dispersion of ppLN, which can be calculated using eq. \eqref{eq:EntanglementTime} and the Sellmeier equations given in reference ~\cite{Gayer2008}. This estimated time is four times shorter than the measured value, but it is only valid for a free-space setup and a bulk crystal. Our large $T_\mathrm{e}$ is disadvantageous for eTPA according to eq. ~\eqref{eq:EntangledCrossSection}.\\

\section{Absorption Measurements}
\label{sec:Absorption Measurements}

\begin{figure*}
    \centering
    \begin{subfigure}{0.32\textwidth}
        \centering
        \includegraphics[width=\textwidth]{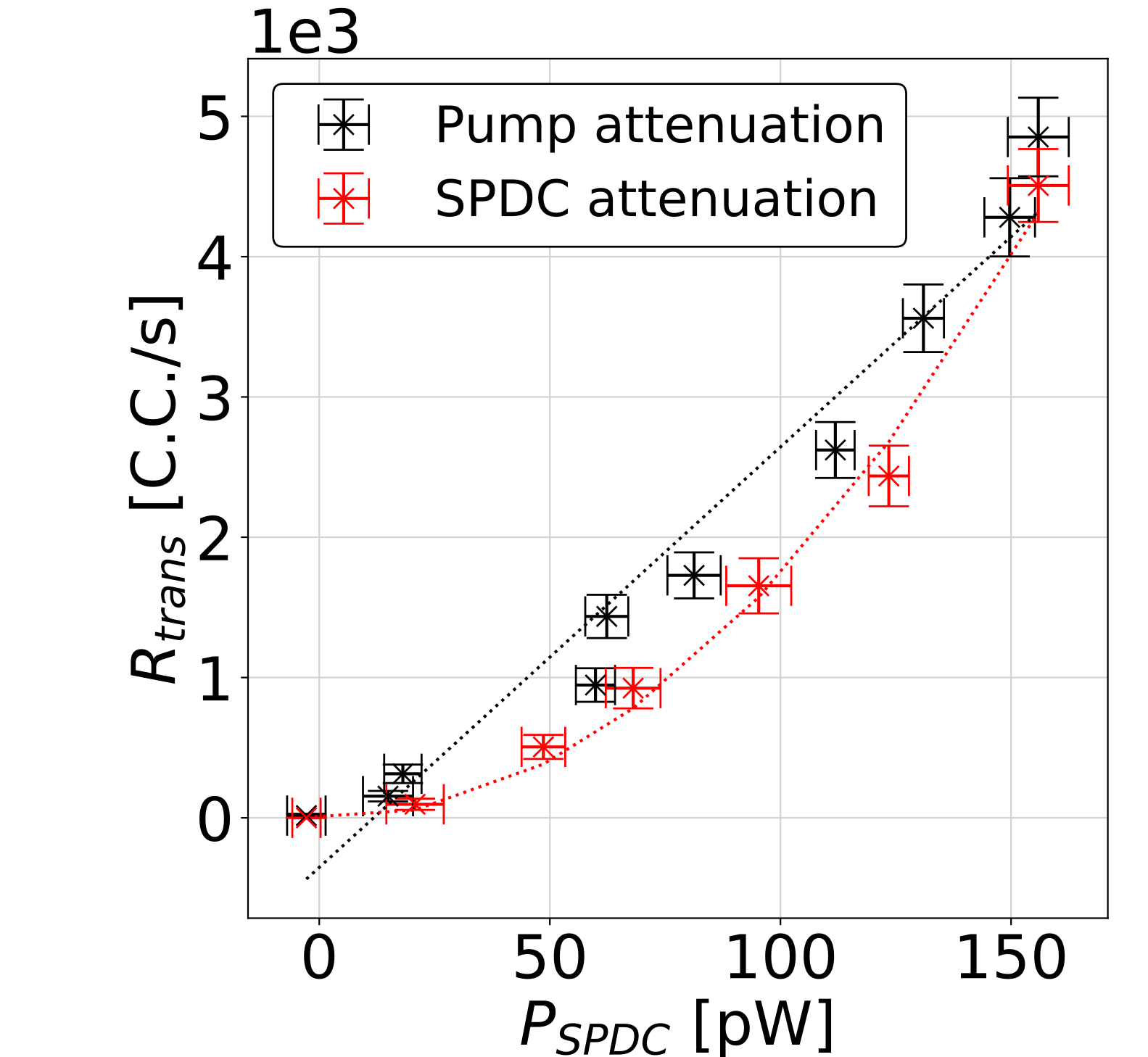}
        \caption{water}
        \label{fig:Rtrans Coin water}
    \end{subfigure}
    \hfill
    \begin{subfigure}{0.32\textwidth}
        \centering
        \includegraphics[width=\textwidth]{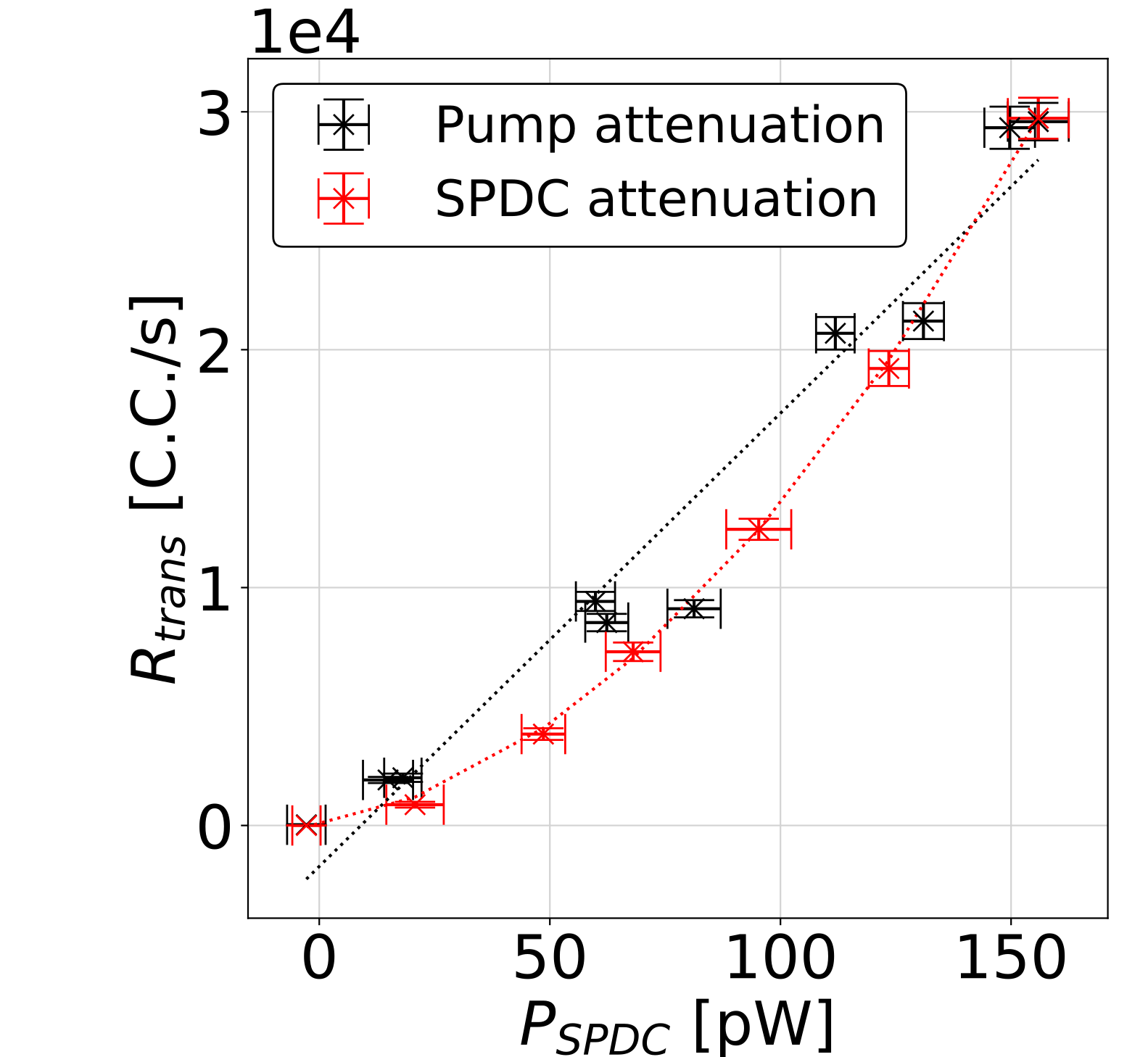}
        \caption{ethanol}
        \label{fig:Rtrans Coin ethanol}
    \end{subfigure}
    \hfill
    \begin{subfigure}{0.32\textwidth}
        \centering
        \includegraphics[width=\textwidth]{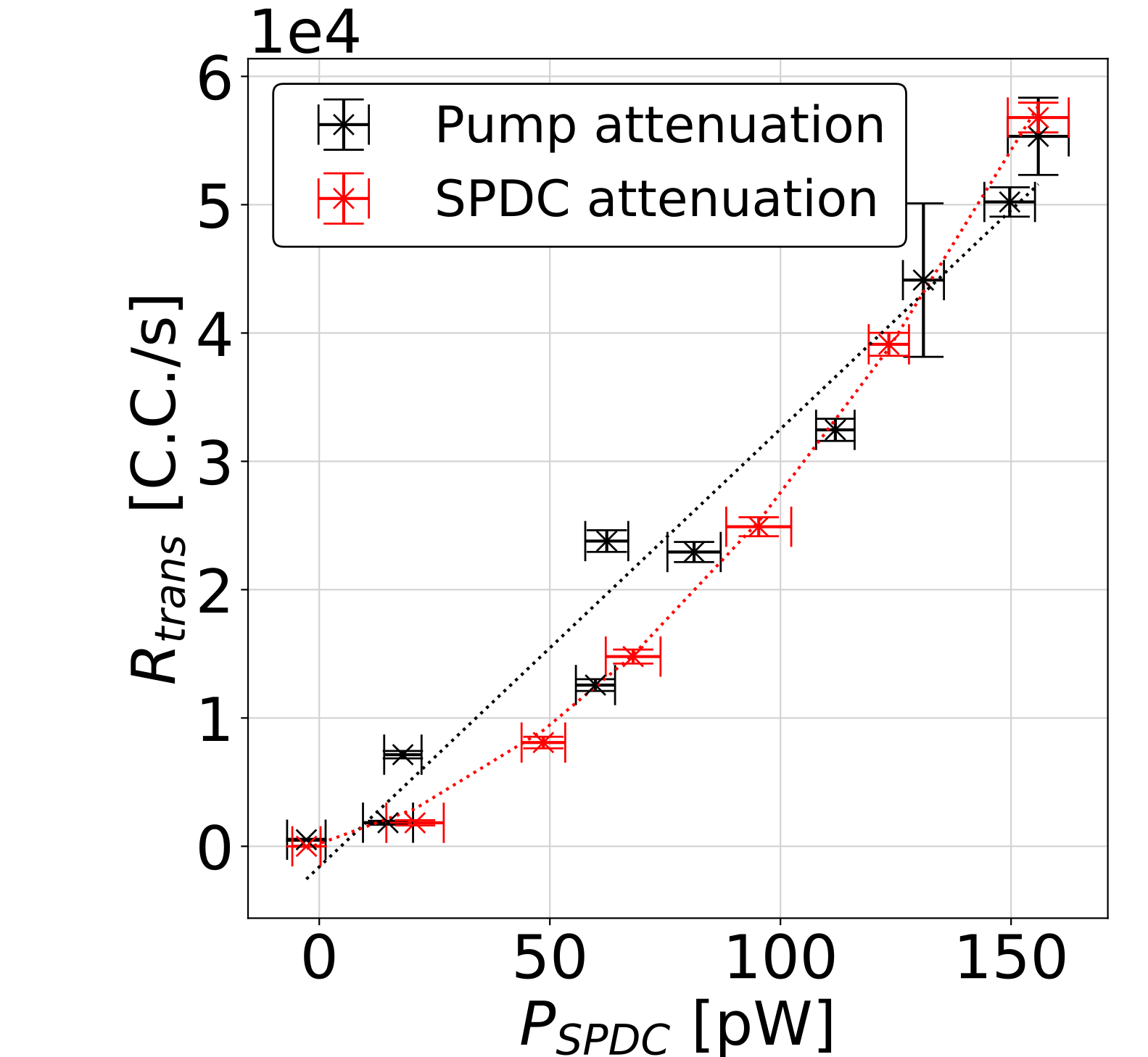}
        \caption{chloroform}
        \label{fig:Rtrans Coin chloroform}
    \end{subfigure}
    \caption{Transmitted coincidence counts $R_{\mathrm{trans}}^{\mathrm{solv}}$ for water (\ref{fig:Rtrans Coin water}), ethanol (\ref{fig:Rtrans Coin ethanol}) and chloroform (\ref{fig:Rtrans Coin chloroform}). Black illustrates the pump attenuation, red the SPDC attenuation.}
    \label{fig:Transmission coincidence Counts solvent}
 \end{figure*}

We have investigated the absorption behavior of different solvents and CdSe/Zns quantum dots by eTPA. For this purpose, the absorption of single photons as well as the reduction of coincidence counts are analysed by a self-made transmission microscope setup (fig. \ref{fig:Setup}b).\\ 
The entangled photon pairs are transmitted from the source to the microscope by a single mode fiber (Thorlabs P1-780AR-2) and decoupled by an adjustable fiber collimator (Thorlabs PAF2P-A15B). This beam is focused down to approximately $\SI{1.7}{\micro\metre}$ onto the sample by a microscope objective (Olympus LUCPLFLN40X). Knowing the numerical aperture $NA$ of this objective, the entanglement area is $A_e\approx\SI{2.13}{\micro\metre^2}$ following the approximation written in eq. \eqref{eq:EntanglementArea}. All investigated samples are filled in cuvettes with $\SI{2}{\milli\metre}$ layer thickness (Hellma 100-2-40). The rates $R_\mathrm{trans}$ of transmitted photon pairs through the sample are collected by an opposite objective (also Olympus LUCPLFLN40X) and fiber-coupled by a large beam collimator (Thorlabs F810FC-780).\\
The detection of these coupled photons is done by the same detector configuration as described in section ~\ref{sec:Pair Generation Efficiency} to maintain the comparability to other publications which already used this configuration \cite{Corona-Aquino2022,Hickam2022a,Martinez-Tapia2023}. Additionally, the pump power will be limited to $P_{\mathrm{pump}}^{\mathrm{max}}\approx\SI{600}{\micro\watt}$ by inserting a ND filter (Thorlabs NE20A-A) directly after the laser to avoid the saturation of the detectors and set the condition that mainly anti-bunched photon pairs arrive at the sample. The generated photon pair rate with the given pump power is approximately $\SI{1.62e7}{\mathrm{C.C./}\second}$. This is maximum available rate for absorption experiments if further losses inside the microscope are neglected. Assuming that this maximum photon pair rate can be focused down to the minimum spot size of our objective, we achieve a photon pair flux of $\phi_{\mathrm{Pair}}^{\mathrm{max}}\approx\SI{7.6e14}{\centi\metre^{-2}\second^{-1}}$.\\
While the detection of the transmission behavior due to the high count of provided photon pairs represents no problem, this setup is not feasible to detect any fluorescence signal excited by the two-photon events. The main reason for that is the low probability of emitting fluorescence photons by eTPA or cTPA which does not surpass the noise level. In addition, light collection by an integrating sphere or other similar collectors inhibits high focusing into the sample, which contradicts the need for small spot sizes for eTPA as apparent from eq. \eqref{eq:EntangledCrossSection} and \eqref{eq:EntanglementArea}. This issue is a crucial point in entangled two-photon microscopy and requires further investigation. However, this work is concentrated on the absorption behavior itself to specify the necessary conditions of fluorescence measurements.\\
In general, absorption measurements with entangled photon pairs show simultaneously linear and quadratic absorption behaviors according to eq. \eqref{eq:AbsorptionRate}. Since the linear term is dominant in the investigated low flux regime, this behavior is removable by the cancellation of entanglement. This cancellation can be done experimentally by placing an absorbing element into the optical path of the entangled photon pairs. Such elements, e.g. neutral density filters, imply a random loss of single photons. Consequently, entanglement between photon pairs is dissolved so that the remaining photons can be absorbed only by cTPA.\\
To investigate the difference between classical and quantum behavior, we measured the transmitted photon counts by reduction of entangled photon pairs via two different ways: First, the pump power are attenuated by the first half-wave plate in the source module, in the following called "pump attenuation". Second, photon pairs generated by maximum pump power are attenuated by inserting different neutral density filters directly after the waveguide, called "SPDC attenuation".\\ 
Fig. \ref{fig:Transmission coincidence Counts solvent} illustrate the photon pair transmission rates $R_{\mathrm{trans}}^{\mathrm{solv}}$ of water, ethanol and chloroform. A regression analysis visualizes that all solvents show a linear behavior for pump attenuation and a weak quadratic behavior for SPDC attenuation. Assuming neglectable absorption or scattering effects inside the solvents, this weak quadratic behavior results mainly from the random losses of single photons caused by the ND filters. The quadratic behavior is supposed to become more significant in the presence of fluorophore molecules due to their absorption of remaining, non-entangled photons by cTPA.

\begin{figure*}
    \centering
    \begin{subfigure}{0.32\textwidth}
        \centering
        \includegraphics[width=\textwidth]{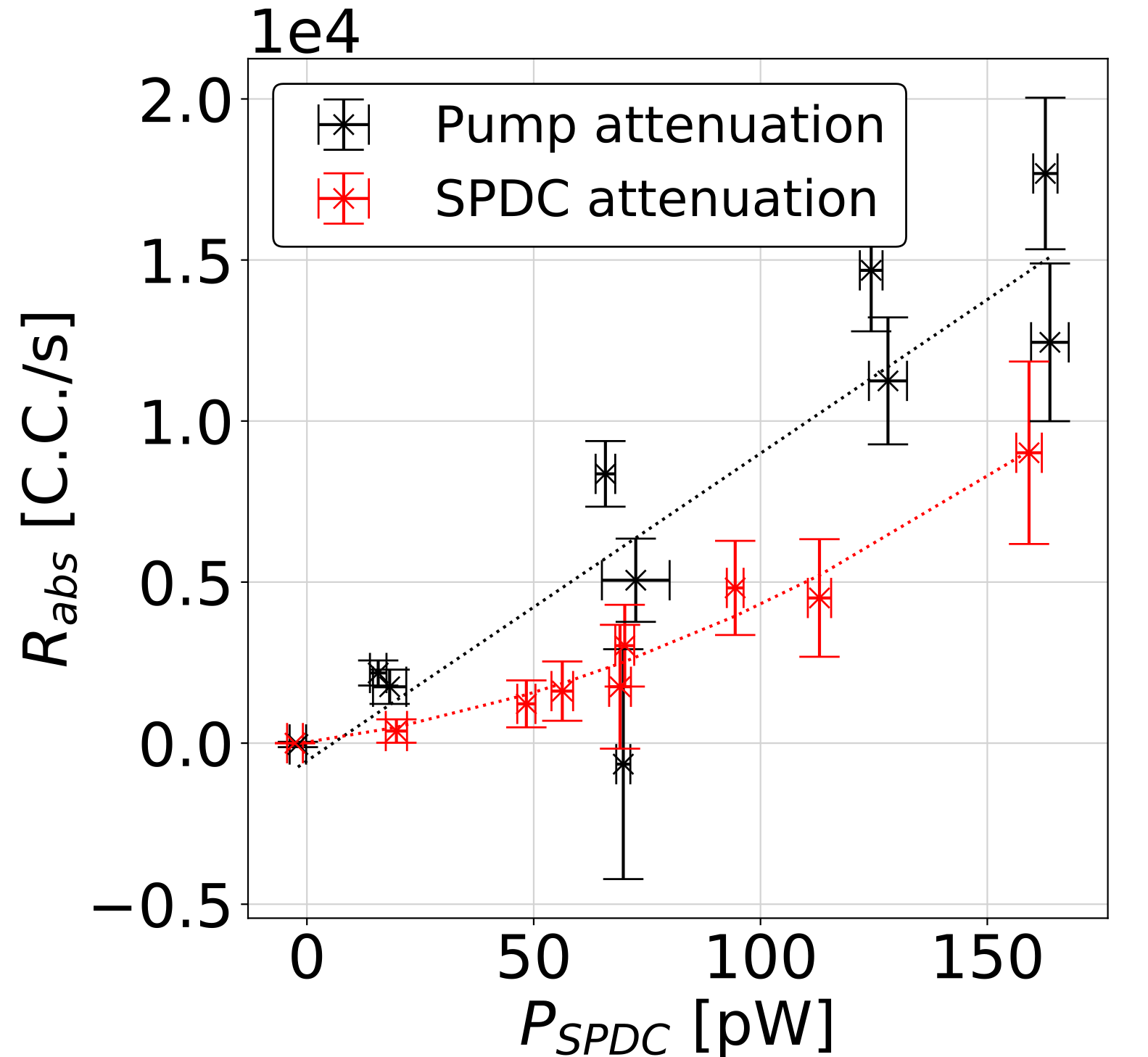}
        \caption{CdSe/ZnS 530}
        \label{fig:Rabs Coin CdSe-ZnS-530}
    \end{subfigure}
    \hfill
    \begin{subfigure}{0.32\textwidth}
        \centering
        \includegraphics[width=\textwidth]{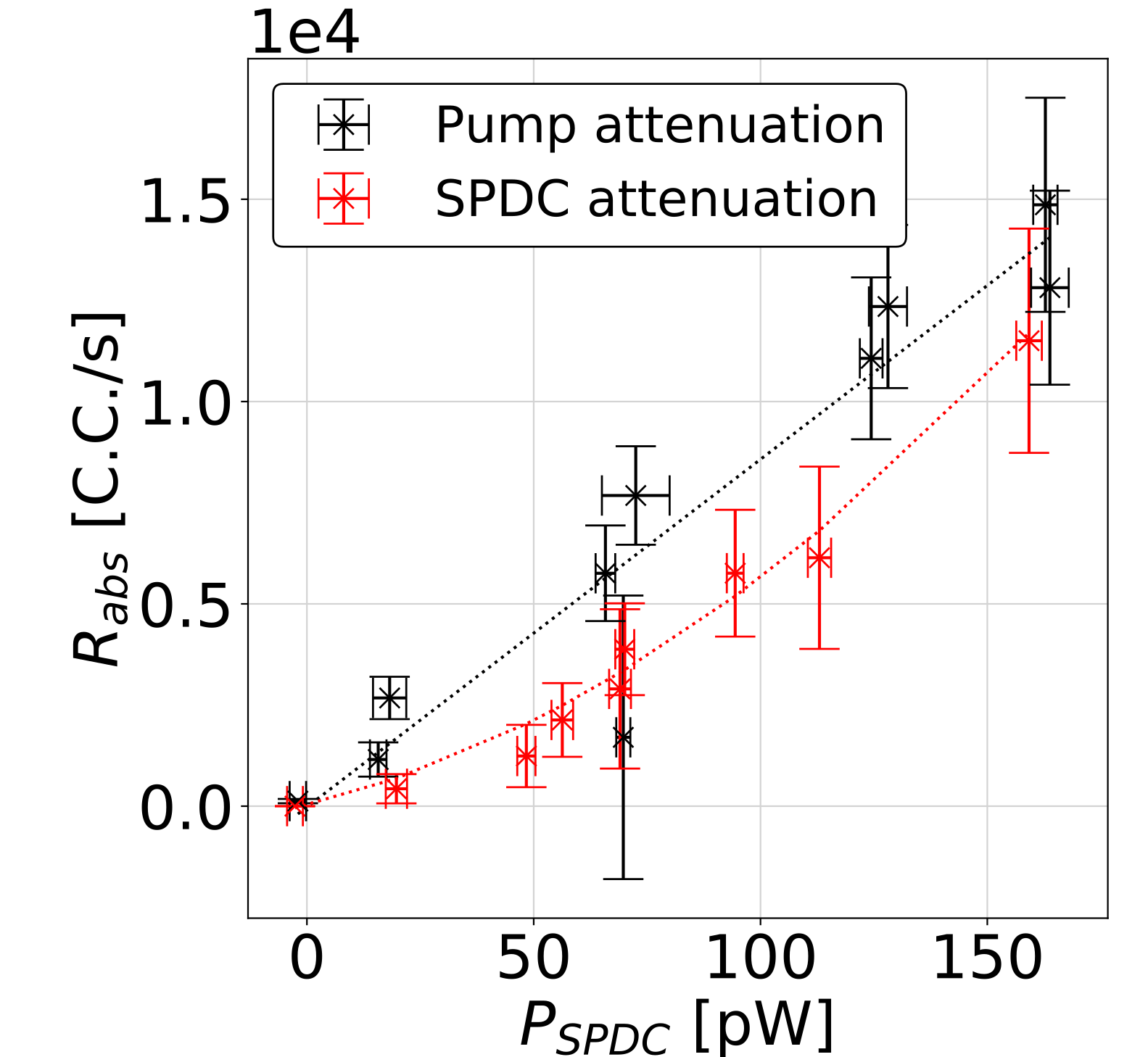}
        \caption{CdSe/ZnS 610}
        \label{fig:Rabs Coin CdSe-ZnS-610}
    \end{subfigure}
    \hfill
    \begin{subfigure}{0.32\textwidth}
        \centering
        \includegraphics[width=\textwidth]{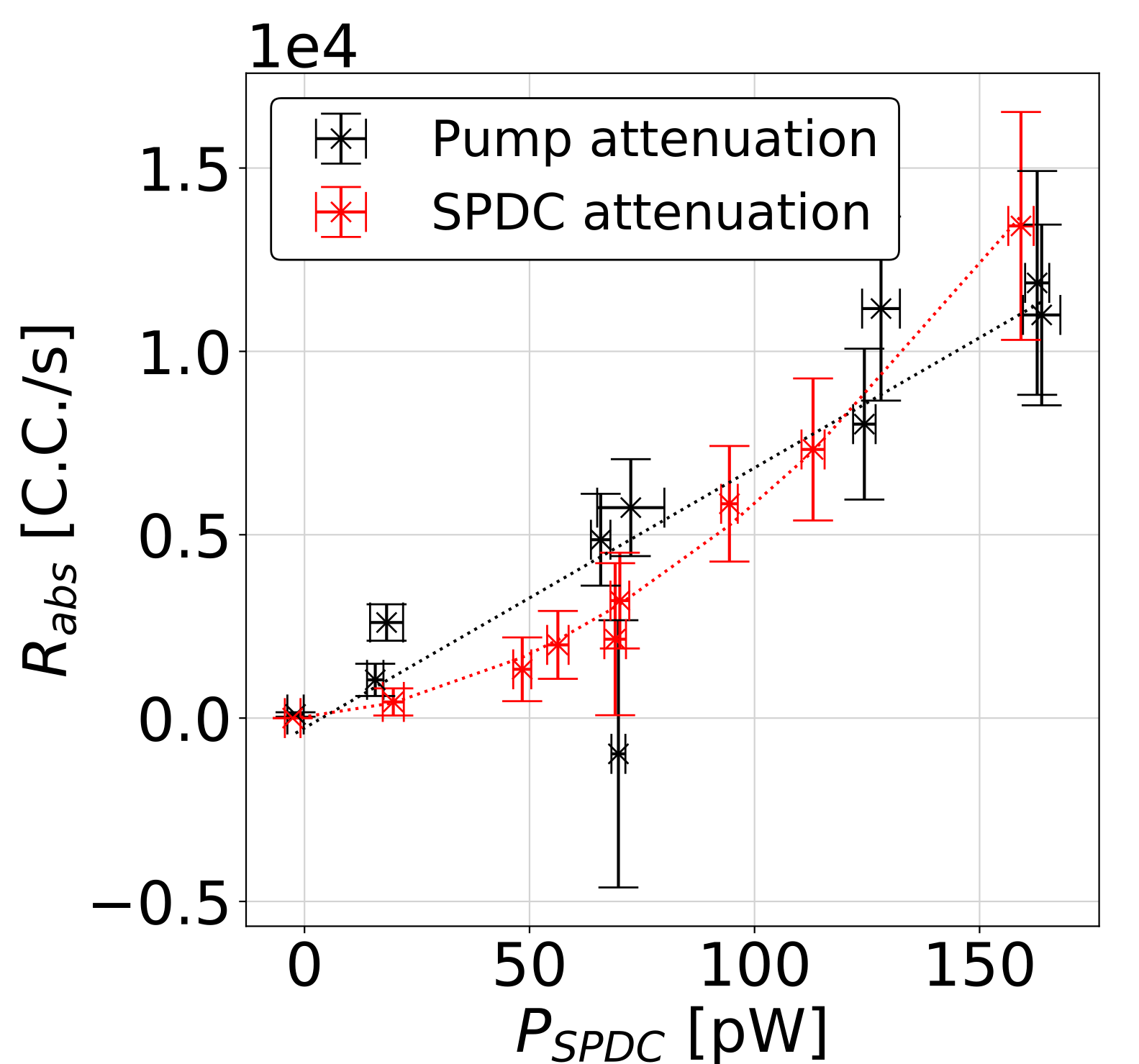}
        \caption{CdSe/ZnS 640}
        \label{fig:Rabs Coin CdSe-ZnS-640}
    \end{subfigure}
    \caption{Absorbed coincidence counts $R_{\mathrm{abs}}$ for quantum dots CdSe/ZnS 530 (\ref{fig:Rabs Coin CdSe-ZnS-530}), CdSe/ZnS 610 (\ref{fig:Rabs Coin CdSe-ZnS-610}) and CdSe/ZnS 640 (\ref{fig:Rabs Coin CdSe-ZnS-640}) solved in chloroform with a concentration of $\SI{1}{\milli\gram/\milli\litre}$. The colors correspond to fig. \ref{fig:Transmission coincidence Counts solvent}.}
    \label{fig:Absorption coincidence Counts probe}
\end{figure*}

However, chloroform (Sigma-Aldrich 288306-100ML) shows the highest amount of transmitted photon pairs. For this purpose, the CdSe/ZnS quantum dots (PlasmaChem PL-QD-O-Kit), which are investigated in the following, are dissolved in chloroform with a mass concentration of $\SI{1}{\milli\gram/\milli\litre}$. This corresponds to a molecule density of about $\SI{2.9e15}{\text{molecules}\per\milli\litre}$ by an average molar mass of $\SI{2.21e5}{\gram/ \mol}$ according to the manufacturer's data.\\ 
Fig. \ref{fig:Absorption coincidence Counts probe} shows the pair absorption rates
\begin{align} 
R_{\mathrm{abs}} &= R_{\mathrm{trans}}^{\mathrm{solv}}-R_{\mathrm{trans}}^{\mathrm{samp}}\label{eq:absorption rate} 
\end{align}
of the CdSe/ZnS quantum dot solutions in which $R_{\mathrm{trans}}^{\mathrm{samp}}$ represents the rate of transmitted pairs through the quantum dot samples. Compared to the transmission behavior of the investigated solvents, the recorded absorption data have a higher uncertainty as visible on the vertical error bars. In particular, the data points at $P_{\mathrm{SPDC}}\approx\SI{160}{\pico\watt}$ for both cases differ significantly. The mismatch for this data point is peculiar because of identical source configuration for pump and SPDC attenuation. The main reason for these discrepancies is to be found in the low amount of absorption events since the measured quantities $R_{\mathrm{trans}}^{\mathrm{solv}}$ and $R_{\mathrm{trans}}^{\mathrm{samp}}$ are in the same order of magnitude. Another noticeable data point is visible at $P_\mathrm{SPDC}\approx\SI{70}{pW}$ for pump attention in all three plots. These points show larger error bars and also differ significant from the general trend. The single count rates for the solvent have relative errors around $\SI{5}{\percent}$ for this point. All other raw data show relative errors much smaller than $\SI{1}{\percent}$. This large errors finally leads to unusually high accidental counts and causes the shown discrepancy. However, an exact reason could not be identified since all measurements were performed consecutively and under the same conditions.

Nevertheless, there is a noticeable difference between pump and SPDC attenuation. But due to the measurement spreads and in accordance with reference ~\cite{Martinez-Tapia2023}, it is not possible to make a clear classification if these differences are caused by eTPA or other parasitic effects, e.g. scattering, hot band absorption ~\cite{Mikhaylov2022} or insufficient detection efficiencies ~\cite{Raymer2021}.\\
For an estimation of expectable absorption rates, the reported classical cross section of $\delta_{\mathrm{c}}\approx\SI{27000}{GM}$ for CdSe/ZnS quantum dots ~\cite{Meiling2018,Parzuchowski2020} and our values for $T_{\mathrm{e}}$ and $A_{\mathrm{e}}$ are used to calculate the entangled cross section of $\sigma_{\mathrm{e}}\approx\SI{3.1e-26}{\centi\metre^2\per\text{molecule}}$ according to eq. ~\eqref{eq:EntangledCrossSection}. Following eq. ~\eqref{eq:AbsorptionRate} this information and the estimated photon pair flux $\phi_{\mathrm{Pair}}^{\mathrm{max}}$ result in absorption rates of $R_{\mathrm{eTPA}}^{\mathrm{max}}\approx\SI{2.4e-11}{\second^{-1}\per\text{molecule}}$ and $R_{\mathrm{cTPA}}^{\mathrm{max}}\approx\SI{1.6e-16}{\second^{-1}\per\text{molecule}}$. Finally, considering an illuminated spherical volume  inside our samples given by the spot size of our objective, the expectable absorption rates are $R_{\mathrm{eTPA}}^{\mathrm{max}}\approx\SI{1.7e-7}{\second^{-1}}$ and $R_{\mathrm{cTPA}}^{\mathrm{max}}\approx\SI{1.1e-12}{\second^{-1}}$.\\
This rough estimation shows that the observation of eTPA is almost impossible with our setup configuration. However, this estimation neglects several elements which affect the probability of eTPA events, for example differing classical cross sections between this study and reference ~\cite{Meiling2018} caused by using another concentration, the enhancement factor in \eqref{eq:EntangledCrossSection} as discussed in reference ~\cite{Landes2021} or the smaller spatial interaction areas for eTPA as shown in reference ~\cite{Tabakaev2022}.\\
Following the concept of reference ~\cite{Corona-Aquino2022}, a more meaningful statement about the presence of eTPA can be given by considering the ratio of biphoton absorption rates $\Gamma$.
\begin{align} 
\Gamma &= 1-\frac{R^\mathrm{Singles1}_\mathrm{samp}R^\mathrm{Singles2}_\mathrm{samp}/R^\mathrm{Coin}_\mathrm{samp}}{R^\mathrm{Singles1}_\mathrm{solv}R^\mathrm{Singles2}_\mathrm{solv}/R^\mathrm{Coin}_\mathrm{solv}} \label{eq:Biphoton ratio}
\end{align}
In eq. \eqref{eq:Biphoton ratio}, $R$ represents the transmitted photon count rates through sample ($\mathrm{samp}$) and solvent ($\mathrm{solv}$). The upper indices of $R$ denote the single counts in channels 1 ($\mathrm{Singles1}$) or 2 ($\mathrm{Singles2}$) and the coincidence rates ($\mathrm{Coin}$). Values of $\Gamma$ different from zero indicate the presence of eTPA, while $\Gamma=0$ demonstrates that obtained losses of photon counts is caused by other effects.

\begin{figure*}
    \centering
    \begin{subfigure}{0.32\textwidth}
        \centering
        \includegraphics[width=\textwidth]{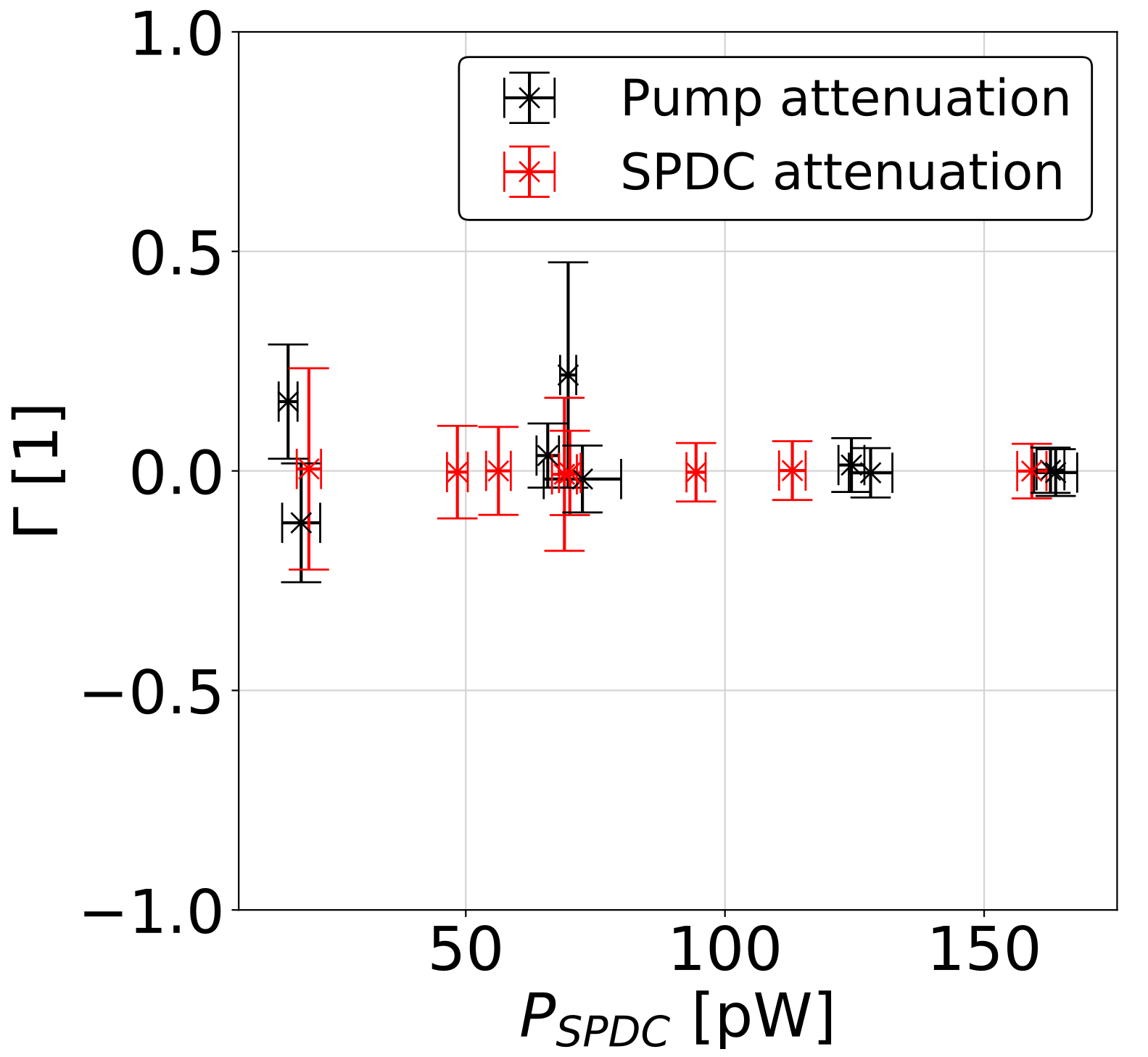}
        \caption{CdSe/ZnS 530}
        \label{fig:Gamma Coin CdSe-ZnS-530}
    \end{subfigure}
    \hfill
    \begin{subfigure}{0.32\textwidth}
        \centering
        \includegraphics[width=\textwidth]{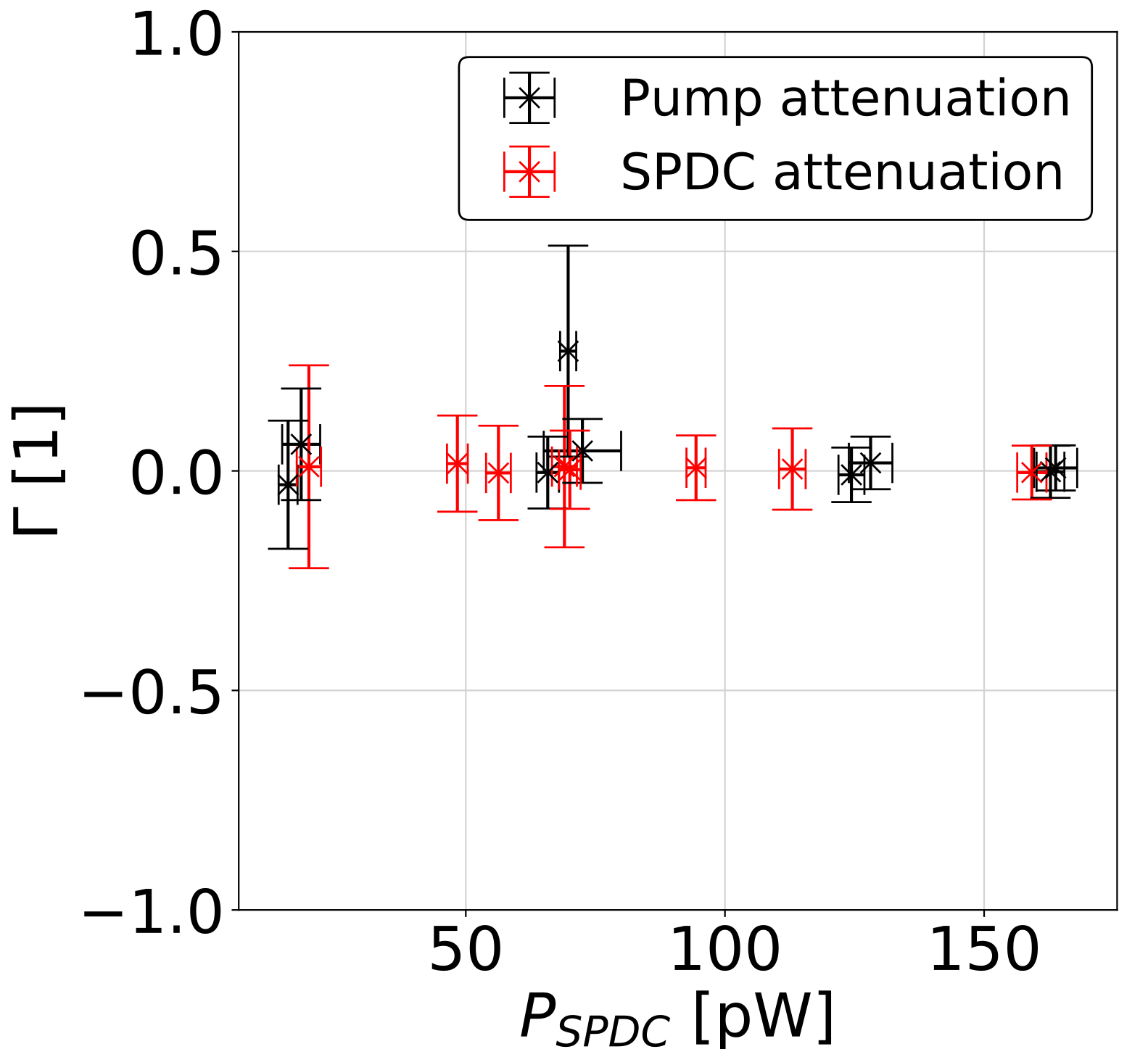}
        \caption{CdSe/ZnS 610}
        \label{fig:Gamma Coin CdSe-ZnS-610}
    \end{subfigure}
    \hfill
    \begin{subfigure}{0.32\textwidth}
        \centering
        \includegraphics[width=\textwidth]{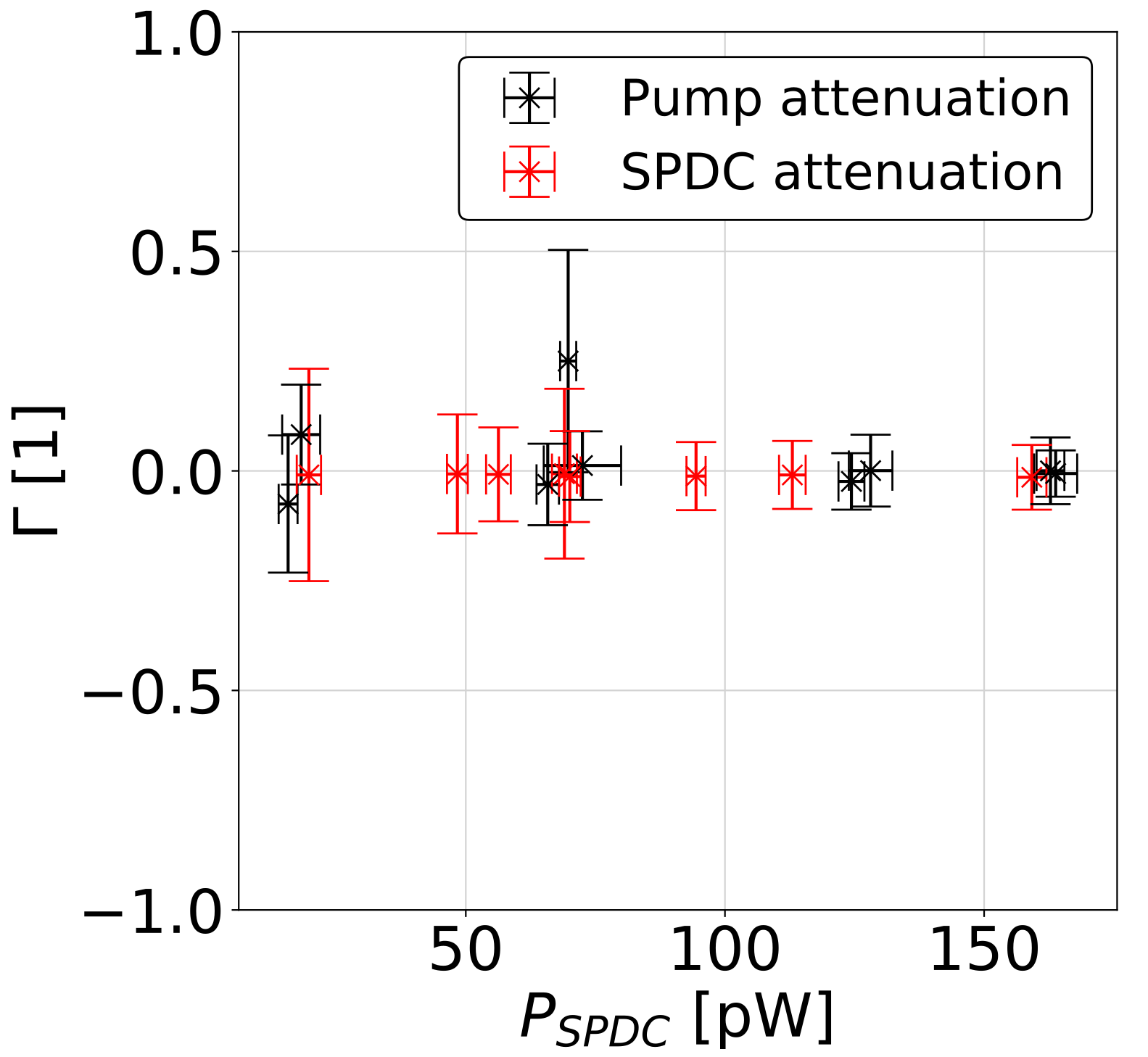}
        \caption{CdSe/ZnS 640}
        \label{fig:Gamma Coin CdSe-ZnS-640}
    \end{subfigure}
    \caption{Biphoton absorption ratio $\Gamma$ for quantum dots CdSe/ZnS 530 (\ref{fig:Gamma Coin CdSe-ZnS-530}), CdSe/ZnS 610 (\ref{fig:Gamma Coin CdSe-ZnS-610}) and CdSe/ZnS 640 (\ref{fig:Gamma Coin CdSe-ZnS-640}) solved in chloroform with a concentration of $\SI{1}{\milli\gram/\milli\litre}$. The colors correspond to fig. \ref{fig:Transmission coincidence Counts solvent}.}
    \label{fig:Biphoton ratio}
\end{figure*}

Fig. ~\ref{fig:Biphoton ratio} shows $\Gamma$ for the investigated quantum dot samples. The data at $P_\mathrm{SPDC}=\SI{0}{pW}$ are not shown since no coincidence counts were detected and which cause large uncertainties in the calculation of eq. \eqref{eq:Biphoton ratio}. All values of $\Gamma$ are close to zero within the error ranges. The reason for the discrepancy of the noticeable data point at $P_\mathrm{SPDC}\approx\SI{70}{pW}$, which also show $\Gamma\neq0$, is probably the unusual high count of single photons measured as already discussed above.\\
Also, these results indicate that the obtained absorption can not be caused by eTPA. To reach a reliable proof of eTPA, more complex detection measurement schemes are needed, for example as recommended in reference ~\cite{Martinez-Tapia2023}, or an unambiguous fluorescence signals.\\

\section{Conclusion}
\label{sec:Conclusion}

We demonstrate that the source of entangled photon pairs based on nonlinear ppLN-waveguides can be a suitable workhorse for fluorescence imaging by eTPA. This study demonstrates that even with the use of a highly efficient photon pair source, the observation of eTPA remains a challenging task, especially when considering the design of the setup for specific applications. However, this technique has the potential for implementation into highly specialized and miniaturized technical developments like optogenetics or lab-on-a-chip devices.\\
This study was restricted to the investigation of CdSe/ZnS quantum dot solutions. To get a larger overview of suitable fluorophores for eTPA, further theoretical and experimental studies are necessary. Also the ambiguous results for the considered absorption rates and non-detection of fluorescence request further improvements of technical realization of fluorescence imaging by eTPA.

\section*{Author Contributions}
\label{sec:Author Contributions}

Conceptualization, T.B.G.; methodology, T.B.G., P.H., N.J. and M.G.; investigation, T.B.G., P.H. and N.J.; data curation, T.B.G.; writing the manuscript, T.B.G., P.H., N.J., and M.G.; supervision, M.G.; project administration, T.B.G. and M.G.; funding acquisition, M.G.\\
All authors have read and agreed to the published version of the manuscript.\\

\section*{Funding}
\label{sec:Funding}
This research was funded by Bundesministerium für Bildung und Forschung (BMBF) through the project "LIVE2QMIC", funding ID 13N15954, within the research focus "Enabling Technologies für die Quantentechnologien".\\

\section*{Data Availability Statement}
\label{sec:Data Availability Statement}

The data presented in this study are available upon request at the corresponding author.\\

\section*{Acknowledgments}
\label{sec:Acknowledgments}

The authors thank Andreas Tünnermann (Fraunhofer Institute of Applied Optics and Precision Engineering IOF Jena and Friedrich Schiller University Jena), Frank Setzpfandt (Friedrich Schiller University Jena), and Christian Eggeling (Leibniz Institute of Photonic Technologies IPHT Jena and Friedrich Schiller University Jena) for fruitful discussions.\\

\section*{Conflicts of Interest}
\label{sec:Conflicts of Interest}

The authors declare no conflict of interest.\\

\bibliography{References.bib}

\end{document}